\documentclass[a4paper, 11pt]{article}
\usepackage[margin=2.5cm]{geometry}
\usepackage{hyperref}

\usepackage[utf8]{inputenc}
\usepackage[english]{babel}
\usepackage{amsthm,amsmath}
\usepackage{hyperref}

\hypersetup{
     colorlinks = true,
     linkcolor = blue!70!black,
     anchorcolor = blue!70!black,
     citecolor = red!70!black,
     filecolor = blue!70!black,
     urlcolor = blue!70!black
}
\usepackage[hyphenbreaks]{breakurl}
\usepackage[all]{hypcap}
\usepackage{stmaryrd}
\usepackage{graphicx}
\usepackage{keycommand}
\usepackage{multicol}
\usepackage{array}

\theoremstyle{definition}
\newtheorem{theorem}{Theorem}[section]

\newtheorem{definition}[theorem]{Definition}

\newtheorem{example}[theorem]{Example}

\newtheorem{example*}[theorem]{Example*}
\newtheorem{examples*}[theorem]{Examples*}

\newtheorem{remark*}[theorem]{Remark*}

\newtheorem*{theorem*}{Theorem}
\newtheorem*{corollary*}{Corollary}
\newtheorem*{lemma*}{Lemma}
\newtheorem*{proposition*}{Proposition}

\usepackage{tikzit}

\tikzstyle{box}=[shape=rectangle, text height=1.5ex, text depth=0.25ex, yshift=0.5mm, fill=white, draw=black, minimum height=5mm, yshift=-0.5mm, minimum width=5mm, font={\small}]
\tikzstyle{gate}=[shape=rectangle, text height=1.5ex, text depth=0.25ex, yshift=0.5mm, fill=white, draw=black, minimum height=5mm, yshift=-0.5mm, minimum width=5mm, font={\small}, tikzit category=circuit]
\tikzstyle{big gate}=[shape=rectangle, text height=1.5ex, text depth=0.25ex, yshift=0.5mm, fill=white, draw=black, minimum height=10mm, yshift=-0.5mm, minimum width=5mm, font={\small}, tikzit category=circuit]
\tikzstyle{Z dot}=[inner sep=0mm, minimum size=2mm, shape=circle, draw=black, fill={rgb,255: red,221; green,255; blue,221}, tikzit category=zx]
\tikzstyle{Z phase dot}=[minimum size=5mm, font={\footnotesize\boldmath}, shape=rectangle, rounded corners=2mm, inner sep=0.2mm, outer sep=-2mm, scale=0.8, tikzit shape=circle, draw=black, fill={rgb,255: red,221; green,255; blue,221}, tikzit draw=blue, tikzit category=zx]
\tikzstyle{X dot}=[Z dot, shape=circle, draw=black, fill={rgb,255: red,255; green,136; blue,136}, tikzit category=zx]
\tikzstyle{X phase dot}=[Z phase dot, tikzit shape=circle, tikzit draw=blue, fill={rgb,255: red,255; green,136; blue,136}, font={\footnotesize\boldmath}, tikzit category=zx]
\tikzstyle{hadamard}=[fill=yellow, draw=black, shape=rectangle, inner sep=0.6mm, minimum height=1.5mm, minimum width=1.5mm, tikzit category=zx]
\tikzstyle{paulibox}=[fill={rgb,255: red,221; green,221; blue,255}, draw=black, shape=rectangle, inner sep=0.6mm, minimum height=5mm, minimum width=5mm, font={\footnotesize}, text height=1.5ex, text depth=0.25ex, tikzit category=zx]
\tikzstyle{vertex}=[inner sep=0mm, minimum size=1mm, shape=circle, draw=black, fill=black, tikzit category=misc]
\tikzstyle{vertex set}=[inner sep=0mm, minimum size=1mm, shape=circle, draw=black, fill=white, font={\footnotesize\boldmath}, tikzit category=misc]
\tikzstyle{small black dot}=[fill=black, draw=black, shape=circle, inner sep=0pt, minimum width=1.2mm, tikzit category=circuit]
\tikzstyle{cnot ctrl}=[fill=black, draw=black, shape=circle, inner sep=0pt, minimum width=1.2mm, tikzit category=circuit]
\tikzstyle{cnot targ}=[fill=white, draw=white, shape=circle, tikzit category=circuit, label={center:$\oplus$}, inner sep=0pt, minimum width=2.1mm, tikzit fill={rgb,255: red,102; green,204; blue,255}, tikzit draw=black]
\tikzstyle{ket}=[fill=white, draw=black, shape=regular polygon, regular polygon sides=3, regular polygon rotate=-30, scale=0.7, inner sep=1pt, tikzit category=circuit, tikzit shape=rectangle, tikzit fill=green]
\tikzstyle{bra}=[fill=white, draw=black, shape=regular polygon, regular polygon sides=3, regular polygon rotate=30, scale=0.7, inner sep=1pt, tikzit category=circuit, tikzit shape=rectangle, tikzit fill=red]
\tikzstyle{scalar}=[shape=rectangle, text height=1.5ex, text depth=0.25ex, yshift=0.5mm, fill=white, draw=black, minimum height=5mm, yshift=-0.5mm, minimum width=5mm, font={\small}]
\tikzstyle{clabel}=[fill=white, draw=none, shape=rectangle, tikzit fill={rgb,255: red,56; green,255; blue,242}, font={\footnotesize}, inner sep=1pt, tikzit category=labels]
\tikzstyle{empty diagram}=[draw={gray!40!white}, dashed, shape=rectangle, minimum width=1cm, minimum height=1cm, tikzit category=misc]

\tikzstyle{simple}=[-]
\tikzstyle{hadamard edge}=[-, dashed, dash pattern=on 2pt off 0.5pt, thick, draw={rgb,255: red,68; green,136; blue,255}]
\tikzstyle{box edge}=[-, dashed, dash pattern=on 2pt off 0.5pt, thick, draw={rgb,255: red,203; green,192; blue,225}]
\tikzstyle{brace edge}=[-, tikzit draw=blue, decorate, decoration={brace,amplitude=1mm,raise=-1mm}]
\tikzstyle{diredge}=[->]
\tikzstyle{double edge}=[-, double, shorten <=-1mm, shorten >=-1mm, double distance=2pt]
\tikzstyle{gray edge}=[-, {gray!60!white}]
\tikzstyle{pointer edge}=[->, very thick, gray]
\tikzstyle{boldedge}=[-, line width=1.6pt, shorten <=-0.17mm, shorten >=-0.17mm]

\input{quantum.tikzdefs}

\usepackage[all]{hypcap} 

\makeatletter
\newcommand\etc{etc\@ifnextchar.{}{.\@}\xspace}

\makeatother

\title{Simulating quantum circuits with ZX-calculus reduced stabiliser decompositions}
\author{Aleks Kissinger and John van de Wetering}

\begin{document}

\maketitle

\begin{abstract}
  We introduce an enhanced technique for strong classical simulation of quantum circuits which combines the `sum-of-stabilisers' method with an automated simplification strategy based on the ZX-calculus.
  Recently it was shown that quantum circuits can be classically simulated by expressing the non-stabiliser gates in a circuit as magic state injections and decomposing them in chunks of 2-6 states at a time, obtaining sums of (efficiently-simulable) stabiliser states with many fewer terms than the naive approach. We adapt these techniques from the original setting of Clifford circuits with magic state injection to generic ZX-diagrams and show that, by interleaving this ``chunked'' decomposition with a ZX-calculus-based simplification strategy, we can obtain stabiliser decompositions that are many orders of magnitude smaller than existing approaches. We illustrate this technique to perform exact norm calculations (and hence strong simulation) on the outputs of random 50- and 100-qubit Clifford+T circuits with up to 70 T-gates as well as a family of hidden shift circuits previously considered by Bravyi and Gosset with over 1000 T-gates.
\end{abstract}

\section{Introduction}

Classical simulation of quantum circuits has a variety of applications, from verifying the correct behaviour of quantum hardware and software to general-purpose simulations of quantum many-body systems. The state of the art in classical simulation also determines whether we can consider a quantum computation to actually give some quantum advantage.
For instance, while Google's quantum supremacy result~\cite{arute2019quantum} was originally believed to take 10,000 years to reproduce on a classical computer, improvements in simulation techniques reduced this to about 20 days~\cite{huangClassicalSimulationQuantum2020} and later to 5 days~\cite{panSimulating2021}.

Classical simulation, or emulation, of quantum computations is widely believed to be a hard problem, requiring exponential classical resources. 
However, there are a variety of different approaches to classical simulation whose time and space requirements vary widely based on the size, shape, contents, or required fidelity of the quantum circuit to be simulated. 
For example, tensor-network based techniques scale exponentially not with the number of qubits, but with the treewidth of the underlying graph of the circuit~\cite{markovSimulatingQuantumComputation2008}. Additionally, noisy quantum circuits are generally easier to simulate~\cite{gaoEfficient2018}.
Recently, methods based on stabiliser rank~\cite{bravyiImprovedClassicalSimulation2016} and stabiliser extent~\cite{Bravyi2019simulationofquantum} have been used to simulate circuits which are suitably close to Clifford circuits.

Thanks to the Gottesman-Knill theorem~\cite{aaronsongottesman2004}, one can efficiently simulate computational basis measurements on the family of quantum stabiliser states, i.e. those states arising from the $\ket{0...0}$ basis state by means of CNOT, $H$, and $S$ gates. It follows immediately from the linearity of the Born rule that one can efficiently simulate any state that is expressed as a linear combination of stabiliser states, as long as the number of terms in the decomposition is not too large. The smallest number of stabiliser terms required to express a given state is known as its \textit{stabiliser rank}. 
While this number is hard to compute in general, heuristics for computing stabiliser decompositions can upper-bound this quantity and can lead to significant speedups in certain families of circuits~\cite{Bravyi2019simulationofquantum}.
In particular, when we have a quantum circuit consisting of Clifford gates and $t$ non-Clifford T gates, then we can represent this as a sum of $2^{\alpha t}$ Clifford circuits for a parameter $\alpha < 1$ which depends on the decomposition strategy. Smaller values of $\alpha$ typically mean we can simulate circuits with larger numbers of T-gates. 
For instance, in~\cite{bravyiImprovedClassicalSimulation2016} they simulated a 40-qubit hidden shift algorithm with about 50 T gates in a couple of hours on a consumer laptop. Their method was improved upon in~\cite{Bravyi2019simulationofquantum} where they could simulate a 50-qubit circuit with 64 T gates.
That work used a decomposition strategy where $\alpha\approx 0.48$, because they decompose a group of 6 T magic states into a sum of 7 stabiliser states, hence obtaining $7^{t/6} \approx 2^{0.48 t}$ terms~\cite{bravyi2016trading}. Using this `chunking' of T magic states, decompositions have been found with $\alpha$ as low (asymptotically) as $0.3963$~\cite{qassim2021improved}.

In this paper we improve upon this technique by combining stabiliser decomposition with a ZX-calculus-based simplification strategy previously applied to circuit optimisation in~\cite{cliffsimp,kissinger2019tcount}. Any quantum circuit, or more generally, any linear map between qubits, can be represented as a special kind of tensor network called a ZX-diagram. The ZX-calculus~\cite{CD1,CD2} then refers to a set of graphical rewrite rules which can be used to transform and simplify ZX-diagrams.

When we translate quantum circuits into ZX-diagrams, the Clifford vs. non-Clifford structure is retained. Namely, Clifford gates become diagrams of \textit{Clifford spiders}, i.e. nodes whose phase parameters are integer multiples of $\frac\pi 2$. We can then treat the remaining non-Clifford spiders analogously to the magic states in Ref.~\cite{bravyiImprovedClassicalSimulation2016} by decomposing them as sums over diagrams of Clifford spiders. Once all non-Clifford spiders are removed, quantum amplitudes can be calculated efficiently by fully simplifying the diagram using a particular rewriting strategy. Furthermore, at each step of the decomposition, we can use this strategy to partially simplify each of the resulting ZX-diagrams as much as possible, enabling non-Clifford spiders to combine and/or cancel out with each other. As we will see, the resulting reduction of non-Clifford spiders can drastically decrease the number of terms in the final stabiliser decomposition.


In this work we focus on Clifford+T circuits, so the non-Clifford spiders are all \textit{T-spiders}, i.e. spiders whose phase angle is an odd multiple of $\frac\pi4$.  Our simplest method, which computes single amplitudes of a circuit, proceeds as follows:
\begin{enumerate}
  \item Write the circuit, together with the desired input state and measurement effect as a ZX-diagram.
  \item Simplify the diagram as much as possible using the ZX-calculus.
  \item Pick a set of the non-Clifford spiders and decompose them, obtaining multiple diagrams with fewer non-Clifford spiders.
  \item Apply the previous two steps recursively to each of the diagrams until no non-Clifford spiders remain, in which case each diagram is simplified to a single complex number.
  \item The sum of these numbers gives the overall amplitude.
\end{enumerate}
Notably, the simplification in step 2 is applied not just to the overall circuit, but eagerly at each step of the decomposition. As we will see in Section~\ref{sec:results}, this can drastically reduce the number of non-Clifford spiders that need to be decomposed, decreasing the size of the sum in step 5 by many orders of magnitude.

Aside from this heuristic benefit, this method also provides a polynomial boost in performance when compared to stabiliser decomposition with standard (i.e. tableu-based) simulation of each of the summands. For calculating a single pure amplitude, our method runs in $O(N^3 + 2^{\alpha t}t^2)$, where $N$ is the total number of gates in the original circuit. Note that this has no dependence on the number of qubits of the circuit. Unless $t$ is very low, this is dominated by the $O(2^{\alpha t}t^2)$ term. In \cite{bravyiImprovedClassicalSimulation2016}, the authors found a corresponding complexity of $O(2^{\alpha t}t^3)$.

Our improvement comes from the fact that we apply simplifications eagerly, so each time we decompose the diagram further, only a constant-sized subdiagram must be further simplified. Using the ZX-calculus, `simulating' a Clifford ZX-diagram amounts to fully simplifying it using the strategy in Section~\ref{sec:simplify-ZX}. Hence, applying simplifications early can be seen as a sort of `partial evaluation' of an almost-Clifford ZX-diagram, where the efficiently-simulable parts are calculated in advance, saving some redundant work later.

Calculating marginal probabilies amounts to computing the norm of a state after some post-selected effects have been applied. To do this, we employ the technique of \emph{doubling} a ZX-diagram~\cite{CKbook,selinger2007dagger}. This doubles the amount of non-Clifford spiders in a diagram, so it could in principle have a catastophic effect on performance. However, we find in practice that many of the additional non-Clifford spiders in this doubled diagram cancel out during simplification, yielding a technique that performs very well even compared to more sophisticated methods for estimating the norm. A pleasant side-effect is that our rewriting-based simulation procedure enables us to compute probabilities exactly as expressions of the form $\frac{1}{2^k}(x + y\sqrt{2})$ for $k, x, y \in \mathbb Z$, as opposed to floating point numbers, which would be subject to rounding errors.

We benchmark our technique on two classes of circuits: random circuits built out of exponentiated Pauli gates and the hidden shift circuits of Ref.~\cite{Bravyi2019simulationofquantum}. We find that we can simulate 50- and 100-qubit exponentiated Pauli circuits with up to 70 T gates, and 50-qubit hidden shift circuits with up to 1400 T gates. This is a significant improvement over the previous largest circuits (in terms of T-count) that have been simulated using the stabiliser decomposition technique. For instance, in Ref.~\cite{Bravyi2019simulationofquantum}, they only simulated a hidden shift circuit with up 64 T gates.

We cover the material that is needed for our paper in Section~\ref{sec:prelim}, in particular, the theory of simulating quantum circuits using stabiliser decompositions, ZX-diagrams, and the ZX-calculus simplification strategy of Ref.~\cite{kissinger2019tcount}.
Then in Section~\ref{sec:method} we describe our simulation technique, and in Section~\ref{sec:results} we present our results when applying this technique to our benchmark circuits. We end with some discussion on possible future improvements to our technique in Section~\ref{sec:conclusion}.

\section{Preliminaries}\label{sec:prelim}

\subsection{Stabiliser decompositions}

The Gottesman-Knill theorem says that we can efficiently simulate a quantum computation as long as the starting state is a stabiliser state, the unitary applied is a Clifford unitary, and the measurement we are doing is a stabiliser measurement~\cite{aaronsongottesman2004}. This holds both for \emph{weak} simulation, where we are merely trying to sample from the output distribution of the quantum computation, as well as for \emph{strong} simulation, where we want to determine (possibly up to some small error) the probability of a particular (marginal) measurement outcome.

Concretely, given a Clifford $n$-qubit unitary $U$ we can determine amplitudes $a_{x_1...x_n} := \bra{x_1\cdots x_n} U \ket{0\cdots 0}$ in $O(n^3)$ time. It immediately follows that we can obtain probabilities of individual measurement outcomes via the Born rule, i.e. $P(x_1 \cdots x_n) = a_{x_1...x_n}^\dagger a_{x_1...x_n}$. 
The Gottesman-Knill theorem also allows us to compute arbitrary marginal probabilies. That is, for $k < n$, we can compute:
\begin{equation}\label{eq:marginal}
  \begin{split}
    P(x_1 \cdots x_k) & = \sum_{x_{k+1}...x_n} P(x_1 \cdots x_n) \\
    & = \sum_{x_{k+1}...x_n} \bra{0\cdots 0}U^\dagger \ket{x_1\cdots x_n}\bra{x_1\cdots x_n} U \ket{0\cdots 0} \\
    & = \bra{0\cdots 0}U^\dagger(\ket{x_1\cdots x_k}\bra{x_1\cdots x_k}\otimes I_{n-k}) U \ket{0\cdots 0}
  \end{split}
\end{equation}

Now suppose we have a Clifford+T circuit. We can then transform it so that every T gate is implemented by a magic-state teleportation and a post-selection, so the computation of an amplitude for this circuit looks like $\bra{x_1\cdots x_n 0\cdots 0}U\ket{0\cdots 0 T\cdots T}$, where the $\ket{0}$'s in the effect correspond to the post-selections, the $U$ is Clifford and $\ket{T}:=\frac{1}{\sqrt{2}}(\ket{0}+e^{i\pi/4}\ket{1})$ is the T magic state. Marginal probabilities can be presented similarly.
The stabiliser states span the entire space of quantum states, and hence we can write this tensor product of T magic states in terms of stabiliser states:
\begin{equation}
\ket{T}^{\otimes n} = \sum_{k=1}^{\chi} \lambda_k \ket{\psi_k}.
\end{equation}
Here $\chi$ denotes the \emph{stabiliser rank} of this decomposition, the amount of stabiliser terms needed to write the state.
With such a \emph{stabiliser decomposition} of the magic states in hand we can calculate the amplitude of the circuit as
\begin{equation}\label{eq:amplitude-magic-decomposition}
\bra{x_1\cdots x_n 0\cdots 0}U\ket{0\cdots 0 T\cdots T} = \sum_{k=1}^{\chi} \lambda_k \bra{x_1\cdots x_n 0\cdots 0}U\ket{0\cdots 0\psi_k}.
\end{equation}
Each of the summands is now just a stabiliser amplitude and hence efficient to calculate. We see then that the efficiency of this method depends on the amount of stabiliser amplitudes we need to calculate, the stabiliser rank.

In general, calculating the minimal stabiliser rank of $\ket{T}^{\otimes n}$ is hard, because the dimension of the search space increases exponentially with $n$. We can however find a (probably not optimal) decomposition by iterating a decomposition for a small number of magic states.
The minimal stabiliser rank of a single T magic state $\ket{T}$ is $2$ as it is not a stabiliser state itself, and has a decomposition $\ket{T} = \frac{1}{\sqrt{2}} \ket{0} + \frac{1}{\sqrt{2}}e^{i\pi/4}\ket{1}$. By taking tensor products of this decomposition we then get a decomposition of $\ket{T}^{\otimes n}$ that has rank $2^n$.
We can however do better.
Note that $\ket{T}\otimes \ket{T} = \frac12(\ket{00}+i\ket{11}) + \frac12e^{i\pi/4}(\ket{01}+\ket{10})$, and hence a pair of T magic states also has stabiliser rank $2$ (rather than 4). Hence, for an even $n$ we can decompose $\ket{T}^{\otimes n}$ into $2^{n/2}$ terms by decomposing the magic states in pairs.
It turns out that there are even more efficient groupings. In Ref.~\cite{bravyi2016trading} it was shown that there is a decomposition of $\ket{T}^{\otimes 6}$ that requires only $7$ terms. Using this decomposition, $n$ magic states require $7^{n/6} = 2^{\alpha n}$ for $\alpha \approx 0.468$ stabiliser terms. This results in better scaling then we would get with the $2^{0.5 n}$ for the pairwise decomposition. By iterating this 6-to-7 decomposition and combining some terms we can get a 12-to-47 decomposition which gives slightly improved exponent of $\alpha \approx 0.463$~\cite{kociaImproved2020}.
Very recently, a new family of decompositions was found, including a 6-to-6 decomposition, which gives an exponent of $\approx 0.4308$. This family of decompositions go all the way down to an asymptotic limit of $\approx 0.3963$~\cite{qassim2021improved}.
Our method relies heavily on a ZX-diagram version of the 6-to-7 decomposition from Ref.~\cite{bravyi2016trading}, whereas we leave incorporating the more recent improved decompositions as future work.

In~\cite{bravyiImprovedClassicalSimulation2016,Bravyi2019simulationofquantum} some techniques are proposed that make the calculation of~\eqref{eq:amplitude-magic-decomposition} faster, for instance by Monte-Carlo sampling weighted by the amplitudes $\lvert \lambda_k\rvert$, as well as enhanced techniques for calculating marginal probabilities using $t$-designs of stabiliser states. We won't consider these methods in this paper, and instead perform strong simulation by exact calculation of the last expression in equation~\eqref{eq:marginal} by stabiliser decomposition.


\subsection{The ZX-calculus}\label{sec:zx}

We will provide a brief overview of the \zxcalculus. For a review see~\cite{vandewetering2020zxcalculus}, and for a book-length introduction see Ref.~\cite{CKbook}.

The \zxcalculus is a diagrammatic language similar to the familiar
quantum circuit notation.  A \emph{\zxdiagram} (or simply
\emph{diagram}) consists of \emph{wires} and \emph{spiders}.  Wires
entering the diagram from the left are \emph{inputs}; wires exiting to
the right are \emph{outputs}.  Given two diagrams we can compose them
by joining the outputs of the first to the inputs of the second, or
form their tensor product by simply stacking the two diagrams.

\emph{Spiders} are linear operations which can have any number of input or output
wires.  There are two varieties: $Z$ spiders depicted as green dots and $X$ spiders depicted as red dots:
\begin{align*}
\small
\begin{tikzpicture}
	\begin{pgfonlayer}{nodelayer}
		\node [style=Z phase dot] (0) at (0, 0) {$\alpha$};
		\node [style=none] (1) at (1.25, 1) {};
		\node [style=none] (2) at (-1.25, 1) {};
		\node [style=none] (3) at (-1.25, -1) {};
		\node [style=none] (4) at (1.25, -1) {};
		\node [style=none] (5) at (1.25, 0.5) {};
		\node [style=none] (6) at (-1.25, 0.5) {};
		\node [style=none, rotate=90] (7) at (-1, -0.25) {...};
		\node [style=none, rotate=90] (8) at (1, -0.25) {...};
	\end{pgfonlayer}
	\begin{pgfonlayer}{edgelayer}
		\draw [in=-141, out=0, looseness=0.75] (3.center) to (0);
		\draw [in=180, out=-39, looseness=0.75] (0) to (4.center);
		\draw [in=180, out=22, looseness=0.75] (0) to (5.center);
		\draw [in=180, out=39, looseness=0.75] (0) to (1.center);
		\draw [in=0, out=158, looseness=0.75] (0) to (6.center);
		\draw [in=141, out=0, looseness=0.75] (2.center) to (0);
	\end{pgfonlayer}
\end{tikzpicture} \ &:= \ \ketbra{0...0}{0...0} +
e^{i \alpha} \ketbra{1...1}{1...1} \\
\begin{tikzpicture}
	\begin{pgfonlayer}{nodelayer}
		\node [style=X phase dot] (0) at (0, 0) {$\alpha$};
		\node [style=none] (1) at (1.25, 1) {};
		\node [style=none] (2) at (-1.25, 1) {};
		\node [style=none] (3) at (-1.25, -1) {};
		\node [style=none] (4) at (1.25, -1) {};
		\node [style=none] (5) at (1.25, 0.5) {};
		\node [style=none] (6) at (-1.25, 0.5) {};
		\node [style=none, rotate=90] (7) at (-1, -0.25) {...};
		\node [style=none, rotate=90] (8) at (1, -0.25) {...};
	\end{pgfonlayer}
	\begin{pgfonlayer}{edgelayer}
		\draw [in=-141, out=0, looseness=0.75] (3.center) to (0);
		\draw [in=180, out=-39, looseness=0.75] (0) to (4.center);
		\draw [in=180, out=22, looseness=0.75] (0) to (5.center);
		\draw [in=180, out=39, looseness=0.75] (0) to (1.center);
		\draw [in=0, out=158, looseness=0.75] (0) to (6.center);
		\draw [in=141, out=0, looseness=0.75] (2.center) to (0);
	\end{pgfonlayer}
\end{tikzpicture} \ &:= \ \ketbra{+...+}{+...+} +
e^{i \alpha} \ketbra{-...-}{-...-}
\end{align*}
Note that if you are reading this document in monochrome or otherwise have difficulty distinguishing green and red, $Z$ spiders will appear lightly-shaded and $X$ darkly-shaded.
The diagram as a whole corresponds to a linear map built from the
spiders (and permutations) by the usual composition and tensor product
of linear maps.  As a special case, diagrams with no inputs represent
(unnormalised) state preparations.

\begin{example}\label{ex:basic-maps-and-states}
  We can immediately write down some simple state preparations and
  unitaries in the \zxcalculus:
  \[
  \begin{array}{rcccl}
  \begin{tikzpicture}
	\begin{pgfonlayer}{nodelayer}
		\node [style=Z dot] (0) at (-0.5, 0) {};
		\node [style=none] (1) at (0.5, 0) {};
	\end{pgfonlayer}
	\begin{pgfonlayer}{edgelayer}
		\draw (0) to (1.center);
	\end{pgfonlayer}
\end{tikzpicture} & = & \ket{0} + \ket{1}& \ =& \sqrt{2}\ket{+} \\[0.2cm]
  \begin{tikzpicture}
	\begin{pgfonlayer}{nodelayer}
		\node [style=X dot] (0) at (-0.5, 0) {};
		\node [style=none] (1) at (0.5, 0) {};
	\end{pgfonlayer}
	\begin{pgfonlayer}{edgelayer}
		\draw (0) to (1.center);
	\end{pgfonlayer}
\end{tikzpicture} & = & \ket{+} + \ket{-}& \ =& \sqrt{2}\ket{0} \\[0.2cm]
  \begin{tikzpicture}
	\begin{pgfonlayer}{nodelayer}
		\node [style=Z phase dot] (0) at (0, 0) {$\alpha$};
		\node [style=none] (1) at (-1.25, 0) {};
		\node [style=none] (2) at (1.25, 0) {};
	\end{pgfonlayer}
	\begin{pgfonlayer}{edgelayer}
		\draw (1.center) to (0);
		\draw (0) to (2.center);
	\end{pgfonlayer}
\end{tikzpicture} & = & \ketbra{0}{0} + e^{i \alpha} \ketbra{1}{1}&\ = & Z_\alpha \\[0.2cm]
  \begin{tikzpicture}
	\begin{pgfonlayer}{nodelayer}
		\node [style=X phase dot] (0) at (0, 0) {$\alpha$};
		\node [style=none] (1) at (-1.25, 0) {};
		\node [style=none] (2) at (1.25, 0) {};
	\end{pgfonlayer}
	\begin{pgfonlayer}{edgelayer}
		\draw (1.center) to (0);
		\draw (0) to (2.center);
	\end{pgfonlayer}
\end{tikzpicture} & = & \ketbra{+}{+} + e^{i \alpha} \ketbra{-}{-}&\ = & X_\alpha
  \end{array}
  \]
  In particular we have the Pauli matrices:
  \[
  \hfill
  \begin{tikzpicture}
	\begin{pgfonlayer}{nodelayer}
		\node [style=Z phase dot] (0) at (0, 0) {$\pi$};
		\node [style=none] (1) at (-1.25, 0) {};
		\node [style=none] (2) at (1.25, 0) {};
	\end{pgfonlayer}
	\begin{pgfonlayer}{edgelayer}
		\draw (1.center) to (0);
		\draw (0) to (2.center);
	\end{pgfonlayer}
\end{tikzpicture} \ \ =\ \  Z \qquad\qquad   \begin{tikzpicture}
	\begin{pgfonlayer}{nodelayer}
		\node [style=X phase dot] (0) at (0, 0) {$\pi$};
		\node [style=none] (1) at (-1.25, 0) {};
		\node [style=none] (2) at (1.25, 0) {};
	\end{pgfonlayer}
	\begin{pgfonlayer}{edgelayer}
		\draw (1.center) to (0);
		\draw (0) to (2.center);
	\end{pgfonlayer}
\end{tikzpicture}\ \ =\ \  X \qquad\qquad
  \hfill
  \]
\end{example}
It will be convenient to introduce a symbol -- a yellow square -- for
the Hadamard gate. This is defined by the equation:
\begin{equation}\label{eq:Hdef}
\hfill
\begin{tikzpicture}
	\begin{pgfonlayer}{nodelayer}
		\node [style=none] (0) at (6.5, 0) {};
		\node [style=none] (1) at (8.5, 0) {};
		\node [style=hadamard] (2) at (7.5, 0) {};
		\node [style=none] (3) at (5.5, 0) {$=:$};
		\node [style=none] (4) at (1.25, 0) {};
		\node [style=Z phase dot] (5) at (2, 0) {$\frac\pi2$};
		\node [style=X phase dot] (6) at (3, 0) {$\frac\pi2$};
		\node [style=Z phase dot] (7) at (4, 0) {$\frac\pi2$};
		\node [style=none] (8) at (4.75, 0) {};
		\node [style=box] (9) at (-3.25, 0) {H};
		\node [style=none] (10) at (-5, 0) {};
		\node [style=none] (11) at (-1.75, 0) {};
		\node [style=none] (12) at (-0.75, 0) {=};
		\node [style=none] (13) at (0.5, 0.075) {$e^{-i\frac\pi4}$};
	\end{pgfonlayer}
	\begin{pgfonlayer}{edgelayer}
		\draw (0.center) to (2);
		\draw (2) to (1.center);
		\draw (4.center) to (8.center);
		\draw (10.center) to (11.center);
	\end{pgfonlayer}
\end{tikzpicture}
\hfill
\end{equation}

We will often use an alternative notation to simplify the diagrams,
and replace a Hadamard between two spiders by a blue dashed edge, as
illustrated below.
\ctikzfig{blue-edge-def} 
Both the blue edge notation and the Hadamard box can always be
translated back into spiders when necessary. We will refer to the blue
edge as a \emph{Hadamard edge}.

Two diagrams are considered \emph{equal} when one can be deformed to
the other by moving the vertices around in the plane, bending,
unbending, crossing, and uncrossing wires, as long as the connectivity
and the order of the inputs and outputs is maintained. Equivalently, a
ZX-diagram can be considered as a graphical depiction of a tensor network,
as in e.g.~\cite{Penrose}. Then, just like any other tensor network, we can observe that the interpretation of a ZX-diagram is unaffected by deformation. As tensors, Z and X spiders can be written as follows:
\begin{align*}
\left( \  \begin{tikzpicture}
	\begin{pgfonlayer}{nodelayer}
		\node [style=Z phase dot] (0) at (0, 0) {$\alpha$};
	\end{pgfonlayer}
\end{tikzpicture}
 \  \right)_{i_1...i_m}^{j_1...j_n} & =
{\small \begin{cases}
1 & \textrm{ if } i_1 = ... = i_m = j_1 = ... = j_n = 0 \\  
e^{i \alpha} & \textrm{ if } i_1 = ... = i_m = j_1 = ... = j_n = 1 \\
0 & \textrm{ otherwise} 
\end{cases}}
\end{align*}
One can then define X-spiders as Z-spiders conjugated by Hadamard gates
or define them explicitly as follows:
\begin{align*}
\left( \  \begin{tikzpicture}
	\begin{pgfonlayer}{nodelayer}
		\node [style=X phase dot] (0) at (0, 0) {$\alpha$};
	\end{pgfonlayer}
\end{tikzpicture}
 \  \right)_{i_1...i_m}^{j_1...j_n} =
    {\small \left(\frac{1}{\sqrt{2}}\right)^{n+m} \cdot 
    \begin{cases}
    1 + e^{i \alpha}\!\! &\!\! \textrm{ if } \bigoplus_\alpha i_\alpha \oplus \bigoplus_\beta j_\beta = 0 \\  
    1 - e^{i \alpha}\!\! &\!\! \textrm{ if } \bigoplus_\alpha i_\alpha \oplus \bigoplus_\beta j_\beta = 1
    \end{cases}}
\end{align*}
where $i_\alpha, j_\beta$ range over $\{0,1\}$ and $\oplus$ is addition modulo~2.

\begin{figure}
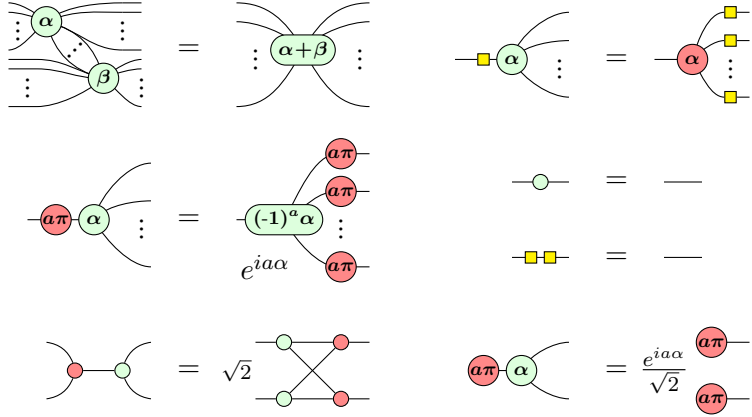

\ctikzfig{ZX-rules}
\caption{\label{fig:zx-rules}
A convenient presentation for the ZX-calculus. These rules hold
  for all $\alpha, \beta \in [0, 2 \pi)$ and $a\in\{0,1\}$. They also hold with the colours (red and green) interchanged, and with the inputs and outputs permuted arbitrarily.}
\end{figure}

In addition to simple deformations, ZX-diagrams satisfy a set of equations called the \zxcalculus. There exists several variations of the ZX-calculus. The set of rules we will use is presented in Figure~\ref{fig:zx-rules}. Diagrams that can be transformed into each other using the rules of the ZX-calculus correspond to equal linear maps. ZX-diagrams with arbitrary angles are expressive enough to represent any linear map~\cite{CD2}. 

We call linear maps that can be produced by combining Clifford unitaries, stabiliser states and stabiliser post-selections \emph{Clifford maps}.
Clifford maps correspond to \textit{Clifford ZX-diagrams}, where all angles are restricted to multiples of $\pi/2$. The rules given in Figure~\ref{fig:zx-rules} are \textit{complete} with respect to linear map equality~\cite{BackensCompleteness} for Clifford ZX-diagrams. That is, if two Clifford ZX-diagrams describe the same linear map, one can be transformed into the other using the rules in Figure~\ref{fig:zx-rules}. Extensions of the calculus exist that are complete for larger families of \zxdiagrams/maps, including \textit{Clifford+T} \zxdiagrams \cite{SimonCompleteness}, where phases are multiples of $\pi/4$, and arbitrary \zxdiagrams where any phase is allowed~\cite{HarnyAmarCompleteness,JPV-universal,vilmarteulercompleteness}.

Quantum circuits can be translated into \zxdiagrams in a straightforward manner. 
We will take as our starting point circuits constructed
from the following set of gates, each of which has a convenient representation in terms of
spiders: 
\begin{align}\label{eq:zx-gates}
\CNOT & = \sqrt{2}\ \begin{tikzpicture}
	\begin{pgfonlayer}{nodelayer}
		\node [style=Z dot] (0) at (0, 0.5) {};
		\node [style=none] (1) at (1.25, 0.5) {};
		\node [style=X dot] (2) at (0, -0.5) {};
		\node [style=none] (3) at (-1.25, 0.5) {};
		\node [style=none] (4) at (1.25, -0.5) {};
		\node [style=none] (5) at (-1.25, -0.5) {};
	\end{pgfonlayer}
	\begin{pgfonlayer}{edgelayer}
		\draw (0) to (2);
		\draw (0) to (3.center);
		\draw (0) to (1.center);
		\draw (5.center) to (2);
		\draw (2) to (4.center);
	\end{pgfonlayer}
\end{tikzpicture} &
Z_\alpha & = \begin{tikzpicture}
	\begin{pgfonlayer}{nodelayer}
		\node [style=Z phase dot] (0) at (0, 0) {$\alpha$};
		\node [style=none] (1) at (-1.25, 0) {};
		\node [style=none] (2) at (1.25, 0) {};
	\end{pgfonlayer}
	\begin{pgfonlayer}{edgelayer}
		\draw (1.center) to (0);
		\draw (0) to (2.center);
	\end{pgfonlayer}
\end{tikzpicture} &
H & = \begin{tikzpicture}
	\begin{pgfonlayer}{nodelayer}
		\node [style=none] (0) at (-0.25, 0) {};
		\node [style=none] (1) at (1.75, 0) {};
		\node [style=hadamard] (2) at (0.75, 0) {};
	\end{pgfonlayer}
	\begin{pgfonlayer}{edgelayer}
		\draw (0.center) to (2);
		\draw (2) to (1.center);
	\end{pgfonlayer}
\end{tikzpicture}
\end{align}
Note that for the CNOT gate, the green spider is the first (i.e.~control) qubit and the red spider is the second (i.e.~target) qubit. 
Other common gates can easily be expressed in terms of these gates. In particular, $S := Z_{\frac\pi2}$, $T := Z_{\frac\pi4}$ and:
\begin{align}\label{eq:zx-derived-gates}
X_\alpha &= \begin{tikzpicture}
	\begin{pgfonlayer}{nodelayer}
		\node [style=X phase dot] (0) at (0, 0) {$\alpha$};
		\node [style=none] (1) at (-1, 0) {};
		\node [style=none] (2) at (1, 0) {};
		\node [style=Z phase dot] (3) at (-4.25, 0) {$\alpha$};
		\node [style=none] (4) at (-5.5, 0) {};
		\node [style=none] (5) at (-3, 0) {};
		\node [style=none] (6) at (-2, 0) {$=$};
		\node [style=hadamard] (7) at (-3.5, 0) {};
		\node [style=hadamard] (8) at (-5, 0) {};
	\end{pgfonlayer}
	\begin{pgfonlayer}{edgelayer}
		\draw (1.center) to (0);
		\draw (0) to (2.center);
		\draw (4.center) to (3);
		\draw (3) to (5.center);
	\end{pgfonlayer}
\end{tikzpicture} \\[0.2cm]
\CZ &\propto \begin{tikzpicture}
	\begin{pgfonlayer}{nodelayer}
		\node [style=Z dot] (0) at (-2.75, 0.5) {};
		\node [style=none] (1) at (-1.5, 0.5) {};
		\node [style=X dot] (2) at (-2.75, -0.5) {};
		\node [style=none] (3) at (-4, 0.5) {};
		\node [style=none] (4) at (-1.5, -0.5) {};
		\node [style=none] (5) at (-4, -0.5) {};
		\node [style=hadamard] (6) at (-3.5, -0.5) {};
		\node [style=hadamard] (7) at (-2, -0.5) {};
		\node [style=none] (8) at (-0.75, 0) {$=$};
		\node [style=Z dot] (9) at (1, 0.5) {};
		\node [style=none] (10) at (2, 0.5) {};
		\node [style=Z dot] (11) at (1, -0.5) {};
		\node [style=none] (12) at (0, 0.5) {};
		\node [style=none] (13) at (2, -0.5) {};
		\node [style=none] (14) at (0, -0.5) {};
		\node [style=hadamard] (15) at (1, 0) {};
		\node [style=none] (16) at (2.75, 0) {$=$};
		\node [style=Z dot] (17) at (4.5, 0.5) {};
		\node [style=none] (18) at (5.5, 0.5) {};
		\node [style=Z dot] (19) at (4.5, -0.5) {};
		\node [style=none] (20) at (3.5, 0.5) {};
		\node [style=none] (21) at (5.5, -0.5) {};
		\node [style=none] (22) at (3.5, -0.5) {};
	\end{pgfonlayer}
	\begin{pgfonlayer}{edgelayer}
		\draw (0) to (2);
		\draw (0) to (3.center);
		\draw (0) to (1.center);
		\draw (5.center) to (2);
		\draw (2) to (4.center);
		\draw (9) to (11);
		\draw (9) to (12.center);
		\draw (9) to (10.center);
		\draw (14.center) to (11);
		\draw (11) to (13.center);
		\draw [style=hadamard edge] (17) to (19);
		\draw (17) to (20.center);
		\draw (17) to (18.center);
		\draw (22.center) to (19);
		\draw (19) to (21.center);
	\end{pgfonlayer}
\end{tikzpicture} \nonumber
\end{align}

\begin{figure}
  \centering
  \begin{tikzpicture}
	\begin{pgfonlayer}{nodelayer}
		\node [style=Z dot] (0) at (-11.5, 2.25) {};
		\node [style=Z phase dot] (2) at (-10.25, 2.25) {$\alpha$};
		\node [style=Z dot] (3) at (-9, 2.25) {};
		\node [style=X dot] (4) at (-11.5, 0.75) {};
		\node [style=X dot] (5) at (-9, 0.75) {};
		\node [style=Z phase dot] (8) at (-7.5, 0.75) {$\gamma$};
		\node [style=none] (10) at (-13.5, 2.25) {};
		\node [style=none] (11) at (-13.5, 0.75) {};
		\node [style=none] (12) at (-6.75, 0.75) {};
		\node [style=none] (13) at (-6.75, 2.25) {};
		\node [style=none] (14) at (-6.75, -0.75) {};
		\node [style=none] (15) at (-13.5, -0.75) {};
		\node [style=hadamard] (173) at (-11.5, -0.75) {};
		\node [style=Z dot] (191) at (-10.25, 0.75) {};
		\node [style=X dot] (192) at (-10.25, -0.75) {};
		\node [style=hadamard] (215) at (-7.75, 2.25) {};
		\node [style=hadamard] (218) at (-12.5, 0.75) {};
		\node [style=none] (219) at (3.5, -7) {$\rightarrow$};
		\node [style=vertex] (220) at (7, -7.75) {};
		\node [style=vertex] (221) at (7, -6.25) {};
		\node [style=vertex] (222) at (5.5, -6.25) {};
		\node [style=vertex] (223) at (8.5, -6.25) {};
		\node [style=vertex] (224) at (7, -4.75) {};
		\node [style=vertex] (225) at (10, -6.25) {};
		\node [style=vertex] (226) at (10, -4.75) {};
		\node [style=vertex] (236) at (10, -7.75) {};
		\node [style=none] (239) at (-13.5, -2.25) {};
		\node [style=none] (240) at (-6.75, -2.25) {};
		\node [style=hadamard] (241) at (-10.25, -2.25) {};
		\node [style=vertex] (247) at (5.5, -9.25) {};
		\node [style=vertex] (248) at (10, -9.25) {};
		\node [style=none] (251) at (-13, -7) {$=$};
		\node [style=Z dot] (252) at (-11, -6.25) {};
		\node [style=Z dot] (253) at (-8, -6.25) {};
		\node [style=Z phase dot] (254) at (-6.5, -6.25) {$\gamma$};
		\node [style=none] (255) at (-12, -4.75) {};
		\node [style=none] (256) at (-12, -6.25) {};
		\node [style=none] (257) at (-5.75, -6.25) {};
		\node [style=none] (258) at (-5.75, -4.75) {};
		\node [style=none] (259) at (-5.75, -7.75) {};
		\node [style=none] (260) at (-12, -7.75) {};
		\node [style=Z dot] (261) at (-9.5, -6.25) {};
		\node [style=Z dot] (262) at (-9.5, -7.75) {};
		\node [style=hadamard] (263) at (-8, -7.75) {};
		\node [style=hadamard] (264) at (-9.5, -7) {};
		\node [style=hadamard] (265) at (-10.25, -6.25) {};
		\node [style=hadamard] (266) at (-8.75, -6.25) {};
		\node [style=hadamard] (267) at (-7.25, -6.25) {};
		\node [style=Z phase dot] (268) at (-9.5, -4.75) {$\alpha$};
		\node [style=hadamard] (269) at (-7.25, -4.75) {};
		\node [style=hadamard, rotate=45] (272) at (-8.75, -5.5) {};
		\node [style=hadamard, rotate=45] (273) at (-10.25, -5.5) {};
		\node [style=none] (274) at (-12, -9.25) {};
		\node [style=none] (275) at (-5.75, -9.25) {};
		\node [style=hadamard] (276) at (-9, -9.25) {};
		\node [style=none] (313) at (-4.75, -7) {$=$};
		\node [style=Z dot] (314) at (-2.75, -6.25) {};
		\node [style=Z dot] (315) at (0.25, -6.25) {};
		\node [style=Z phase dot] (316) at (1.75, -6.25) {$\gamma$};
		\node [style=none] (317) at (-3.75, -4.75) {};
		\node [style=none] (318) at (-3.75, -6.25) {};
		\node [style=none] (319) at (2.5, -6.25) {};
		\node [style=none] (320) at (2.5, -4.75) {};
		\node [style=none] (321) at (2.5, -7.75) {};
		\node [style=none] (322) at (-3.75, -7.75) {};
		\node [style=Z dot] (323) at (-1.25, -6.25) {};
		\node [style=Z dot] (324) at (-1.25, -7.75) {};
		\node [style=hadamard] (325) at (0.25, -7.75) {};
		\node [style=hadamard] (326) at (-1.25, -7) {};
		\node [style=hadamard] (327) at (-2, -6.25) {};
		\node [style=hadamard] (328) at (-0.5, -6.25) {};
		\node [style=hadamard] (329) at (1, -6.25) {};
		\node [style=Z phase dot] (330) at (-1.25, -4.75) {$\alpha$};
		\node [style=hadamard] (331) at (1, -4.75) {};
		\node [style=Z dot] (332) at (1.75, -4.75) {};
		\node [style=Z dot] (333) at (1.75, -7.75) {};
		\node [style=hadamard, rotate=45] (334) at (-0.5, -5.5) {};
		\node [style=hadamard, rotate=45] (335) at (-2, -5.5) {};
		\node [style=none] (336) at (-3.75, -9.25) {};
		\node [style=none] (337) at (2.5, -9.25) {};
		\node [style=hadamard] (338) at (-0.5, -9.25) {};
		\node [style=Z dot] (339) at (-2.75, -9.25) {};
		\node [style=Z dot] (340) at (1.75, -9.25) {};
		\node [style=Z dot] (346) at (-2.75, 2.25) {};
		\node [style=Z phase dot] (347) at (-1.5, 2.25) {$\alpha$};
		\node [style=Z dot] (348) at (-0.25, 2.25) {};
		\node [style=Z dot] (349) at (-2.75, 0.75) {};
		\node [style=Z dot] (350) at (-0.25, 0.75) {};
		\node [style=Z phase dot] (351) at (1.25, 0.75) {$\gamma$};
		\node [style=none] (352) at (-4.75, 2.25) {};
		\node [style=none] (353) at (-4.75, 0.75) {};
		\node [style=none] (354) at (2, 0.75) {};
		\node [style=none] (355) at (2, 2.25) {};
		\node [style=none] (356) at (2, -0.75) {};
		\node [style=none] (357) at (-4.75, -0.75) {};
		\node [style=hadamard] (358) at (-3.25, -0.75) {};
		\node [style=Z dot] (359) at (-1.5, 0.75) {};
		\node [style=Z dot] (360) at (-1.5, -0.75) {};
		\node [style=hadamard] (361) at (1, 2.25) {};
		\node [style=hadamard] (362) at (-4.25, 0.75) {};
		\node [style=none] (363) at (-4.75, -2.25) {};
		\node [style=none] (364) at (2, -2.25) {};
		\node [style=hadamard] (365) at (-1.5, -2.25) {};
		\node [style=none] (366) at (-5.75, -0.25) {$=$};
		\node [style=hadamard] (368) at (-3.5, 0.75) {};
		\node [style=hadamard] (369) at (-2.125, 0.75) {};
		\node [style=hadamard] (370) at (-0.875, 0.75) {};
		\node [style=hadamard] (371) at (0.375, 0.75) {};
		\node [style=hadamard] (372) at (-2.375, -0.75) {};
		\node [style=hadamard] (373) at (-0.5, -0.75) {};
		\node [style=hadamard] (374) at (-2.75, 1.5) {};
		\node [style=hadamard] (375) at (-0.25, 1.5) {};
		\node [style=hadamard] (376) at (-1.5, 0) {};
		\node [style=none] (377) at (3, -0.25) {$=$};
		\node [style=Z dot] (379) at (6, 2.25) {};
		\node [style=Z phase dot] (380) at (7.25, 2.25) {$\alpha$};
		\node [style=Z dot] (381) at (8.5, 2.25) {};
		\node [style=Z dot] (382) at (6, 0.75) {};
		\node [style=Z dot] (383) at (8.5, 0.75) {};
		\node [style=Z phase dot] (384) at (10, 0.75) {$\gamma$};
		\node [style=none] (385) at (4, 2.25) {};
		\node [style=none] (386) at (4, 0.75) {};
		\node [style=none] (387) at (10.75, 0.75) {};
		\node [style=none] (388) at (10.75, 2.25) {};
		\node [style=none] (389) at (10.75, -0.75) {};
		\node [style=none] (390) at (4, -0.75) {};
		\node [style=Z dot] (392) at (7.25, 0.75) {};
		\node [style=Z dot] (393) at (7.25, -0.75) {};
		\node [style=hadamard] (394) at (9.75, 2.25) {};
		\node [style=none] (396) at (4, -2.25) {};
		\node [style=none] (397) at (10.75, -2.25) {};
		\node [style=hadamard] (398) at (7.25, -2.25) {};
		\node [style=hadamard] (400) at (6.625, 0.75) {};
		\node [style=hadamard] (401) at (7.875, 0.75) {};
		\node [style=hadamard] (402) at (9.125, 0.75) {};
		\node [style=hadamard] (404) at (8.25, -0.75) {};
		\node [style=hadamard] (405) at (6, 1.5) {};
		\node [style=hadamard] (406) at (8.5, 1.5) {};
		\node [style=hadamard] (407) at (7.25, 0) {};
	\end{pgfonlayer}
	\begin{pgfonlayer}{edgelayer}
		\draw (0) to (4);
		\draw (2) to (3);
		\draw (3) to (5);
		\draw (5) to (8);
		\draw (10.center) to (0);
		\draw (11.center) to (4);
		\draw (4) to (5);
		\draw (8) to (12.center);
		\draw (0) to (2);
		\draw (15.center) to (173);
		\draw (173) to (14.center);
		\draw (191) to (192);
		\draw (3) to (13.center);
		\draw (224) to (226);
		\draw (224) to (223);
		\draw (222) to (224);
		\draw (220) to (221);
		\draw (222) to (221);
		\draw (221) to (223);
		\draw (223) to (225);
		\draw (236) to (220);
		\draw (239.center) to (240.center);
		\draw (247) to (248);
		\draw (253) to (254);
		\draw (256.center) to (252);
		\draw (252) to (253);
		\draw (254) to (257.center);
		\draw (261) to (262);
		\draw (260.center) to (259.center);
		\draw [loop] (268) to ();
		\draw (268) to (253);
		\draw (268) to (252);
		\draw (255.center) to (268);
		\draw [loop] (268) to ();
		\draw (268) to (258.center);
		\draw (274.center) to (275.center);
		\draw (315) to (316);
		\draw (318.center) to (314);
		\draw (314) to (315);
		\draw (316) to (319.center);
		\draw (323) to (324);
		\draw (322.center) to (321.center);
		\draw [loop] (330) to ();
		\draw (330) to (315);
		\draw (330) to (314);
		\draw (317.center) to (330);
		\draw [loop] (330) to ();
		\draw (330) to (320.center);
		\draw (336.center) to (337.center);
		\draw (346) to (349);
		\draw (347) to (348);
		\draw (348) to (350);
		\draw (350) to (351);
		\draw (352.center) to (346);
		\draw (353.center) to (349);
		\draw (349) to (350);
		\draw (351) to (354.center);
		\draw (346) to (347);
		\draw (357.center) to (358);
		\draw (358) to (356.center);
		\draw (359) to (360);
		\draw (348) to (355.center);
		\draw (363.center) to (364.center);
		\draw (379) to (382);
		\draw (380) to (381);
		\draw (381) to (383);
		\draw (383) to (384);
		\draw (385.center) to (379);
		\draw (386.center) to (382);
		\draw (382) to (383);
		\draw (384) to (387.center);
		\draw (379) to (380);
		\draw (392) to (393);
		\draw (381) to (388.center);
		\draw (396.center) to (397.center);
		\draw (390.center) to (389.center);
	\end{pgfonlayer}
\end{tikzpicture}
  \caption{\label{fig:graph-like} A \zxdiagram that comes from a circuit, and its reduction to an equivalent graph-like \zxdiagram, and the corresponding underlying simple graph.}
\end{figure}
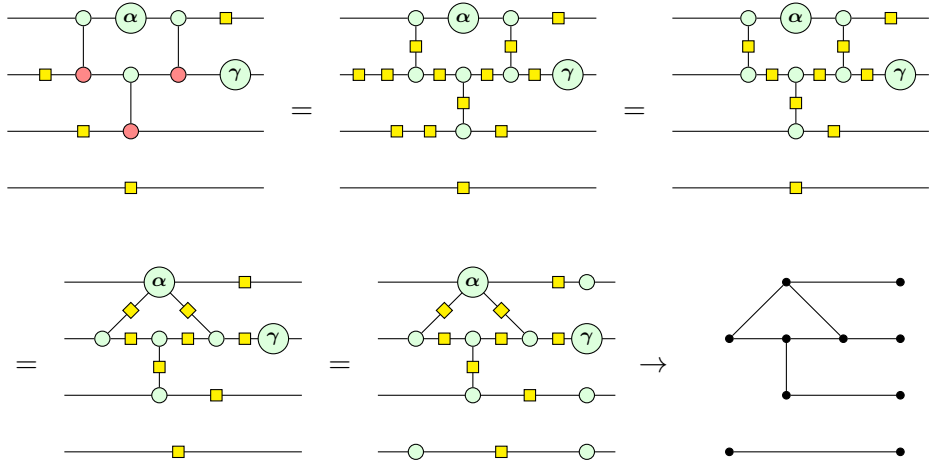

For purposes of automated simplification of ZX-diagrams it is often useful to transform the diagram into a \emph{graph-like} \zxdiagram~\cite{cliffsimp}.

\begin{definition}\label{def:graph-form}
  A \zxdiagram is \emph{graph-like} when:
  \begin{enumerate}
    \item All spiders are Z-spiders.
    \item Z-spiders are only connected via Hadamard edges.
    \item There are no parallel Hadamard edges or self-loops.
    \item Every input or output is connected to a Z-spider and every Z-spider is connected to at most one input or output.
  \end{enumerate}
\end{definition}

In Ref.~\cite{cliffsimp} it is shown that any \zxdiagram can efficiently be transformed into a graph-like \zxdiagram using the rules of the ZX-calculus. This transformation essentially amounts to turning all X spiders into Z spiders with the colour-change rule (top-right in Fig.~\ref{fig:zx-rules}), fusing as many spiders together as possible, and removing parallel edges/self-loops with the following derived rules:
\begin{align}\label{eq:parallel-edges-loops}
\begin{tikzpicture}
	\begin{pgfonlayer}{nodelayer}
		\node [style=Z phase dot] (0) at (-4.75, 0) {$\alpha$};
		\node [style=Z phase dot] (1) at (-2.25, 0) {$\beta$};
		\node [style=none] (2) at (-6, 1) {};
		\node [style=none] (3) at (-6, -1) {};
		\node [style=none] (4) at (-1, 1) {};
		\node [style=none] (5) at (-1, -1) {};
		\node [style=none] (6) at (-6, 0) {...};
		\node [style=none] (7) at (-1, 0) {...};
		\node [style=none] (10) at (0, 0) {=};
		\node [style=Z phase dot] (11) at (3.25, 0) {$\alpha$};
		\node [style=Z phase dot] (12) at (4.75, 0) {$\beta$};
		\node [style=none] (13) at (2, 1) {};
		\node [style=none] (14) at (2, -1) {};
		\node [style=none] (15) at (6, 1) {};
		\node [style=none] (16) at (6, -1) {};
		\node [style=none] (17) at (2, 0) {...};
		\node [style=none] (18) at (6, 0) {...};
		\node [style=none] (19) at (1, 0) {$\frac12$};
	\end{pgfonlayer}
	\begin{pgfonlayer}{edgelayer}
		\draw (0) to (2.center);
		\draw (0) to (3.center);
		\draw (1) to (4.center);
		\draw (1) to (5.center);
		\draw (11) to (13.center);
		\draw (11) to (14.center);
		\draw (12) to (15.center);
		\draw (12) to (16.center);
		\draw [style=hadamard edge, bend right] (0) to (1);
		\draw [style=hadamard edge, bend left] (0) to (1);
	\end{pgfonlayer}
\end{tikzpicture} \\
\begin{tikzpicture}
	\begin{pgfonlayer}{nodelayer}
		\node [style=Z phase dot] (0) at (-1.75, 0) {$\alpha$};
		\node [style=none] (1) at (-2.75, -0.75) {};
		\node [style=none] (2) at (-0.75, -0.75) {};
		\node [style=none] (3) at (-1.75, 1) {};
		\node [style=none] (4) at (-1.75, -0.75) {...};
		\node [style=none] (5) at (0, 0) {=};
		\node [style=none] (6) at (1.5, -0.75) {...};
		\node [style=none] (7) at (0.5, -0.75) {};
		\node [style=none] (8) at (2.5, -0.75) {};
		\node [style=Z phase dot] (9) at (1.5, 0) {$\alpha$};
	\end{pgfonlayer}
	\begin{pgfonlayer}{edgelayer}
		\draw (1.center) to (0);
		\draw (0) to (2.center);
		\draw [in=360, out=15, looseness=1.50] (0) to (3.center);
		\draw [in=540, out=165, looseness=1.50] (0) to (3.center);
		\draw (7.center) to (9);
		\draw (9) to (8.center);
	\end{pgfonlayer}
\end{tikzpicture} \qquad\quad\ \ \  {}\nonumber\\
\begin{tikzpicture}
	\begin{pgfonlayer}{nodelayer}
		\node [style=Z phase dot] (0) at (-1.75, 0) {$\alpha$};
		\node [style=none] (1) at (-2.75, -0.75) {};
		\node [style=none] (2) at (-0.75, -0.75) {};
		\node [style=none] (3) at (-1.75, 1) {};
		\node [style=none] (4) at (-1.75, -0.75) {...};
		\node [style=none] (5) at (0, 0) {=};
		\node [style=none] (6) at (2.5, -0.75) {...};
		\node [style=none] (7) at (1.5, -0.75) {};
		\node [style=none] (8) at (3.5, -0.75) {};
		\node [style=Z phase dot] (9) at (2.5, 0) {$\ \alpha + \pi\ $};
		\node [style=none] (10) at (1, 0) {$\frac{1}{\sqrt{2}}$};
	\end{pgfonlayer}
	\begin{pgfonlayer}{edgelayer}
		\draw (1.center) to (0);
		\draw (0) to (2.center);
		\draw [style=hadamard edge, in=360, out=15, looseness=1.50] (0) to (3.center);
		\draw [style=hadamard edge, in=540, out=165, looseness=1.50] (0) to (3.center);
		\draw (7.center) to (9);
		\draw (9) to (8.center);
	\end{pgfonlayer}
\end{tikzpicture} \qquad\quad \ \   {}\nonumber
\end{align}
See Fig.~\ref{fig:graph-like} for an example.
Note that this procedure never increases the number of non-Clifford phases, but that it can actually decrease the number by fusing spiders, as phases are added together.
We call these diagrams \textit{graph-like} because the resulting \zxdiagram is essentially an indirected, simple graph whose vertices are labelled by phase angles.

\subsection{Simplifying ZX-diagrams}\label{sec:simplify-ZX}

In this paper we use the simplification strategy for ZX-diagrams described in~\cite{kissinger2019tcount}. We will give a brief summary of this method here.
Note that all of the rules in this section can be derived from the basic rules of the ZX-calculus given in Figure~\ref{fig:zx-rules}. For the details see~\cite{kissinger2019tcount}.

The main idea of this simplification strategy is that we want to remove as many spiders from the diagram as possible. The first step is that we reduce the diagram to graph-like form, so that all the spiders are Z-spiders, and the only connectivity is via Hadamard edges.
The main tools for the simplification are two simplification rules called \emph{local complementation} and \emph{pivoting}. The first of these removes an \emph{internal} spider (one who is not connected to an input or output wire) with a $\pm\frac\pi2$ phase, at the cost of complementing the connectivity in its neighbourhood and changing some phases:
\begin{equation}\label{eq:lc-simp}
  \begin{tikzpicture}
	\begin{pgfonlayer}{nodelayer}
		\node [style=Z phase dot] (0) at (-5, 0) {$\ \pm\frac\pi2\ $};
		\node [style=Z phase dot] (2) at (-7.5, -1.25) {$\alpha_1$};
		\node [style=Z phase dot] (4) at (-2.25, -1.25) {$\alpha_n$};
		\node [style=none] (5) at (-8.5, -2.25) {};
		\node [style=none] (6) at (-7.5, -2.25) {};
		\node [style=none] (7) at (-2.25, -2.25) {};
		\node [style=none] (8) at (-1.25, -2.25) {};
		\node [style=none] (9) at (-5, -2) {...};
		\node [style=none] (10) at (-8, -2.25) {...};
		\node [style=none] (11) at (-1.75, -2.25) {...};
		\node [style=none] (16) at (0, -1.5) {$=$};
		\node [style=none] (17) at (15.25, -1.25) {};
		\node [style=none] (18) at (7, -1.25) {};
		\node [style=none] (19) at (14.75, -1.25) {...};
		\node [style=none] (20) at (14.25, -1.25) {};
		\node [style=Z phase dot] (24) at (8, -0.25) {$\ \alpha_1\!\mp\!\frac\pi2\ $};
		\node [style=none] (26) at (8, -1.25) {};
		\node [style=none] (27) at (7.5, -1.25) {...};
		\node [style=Z phase dot] (30) at (14.25, -0.25) {$\ \alpha_n \!\mp\! \frac\pi2\ $};
		\node [style=none] (31) at (-6.5, -4.25) {};
		\node [style=Z phase dot] (32) at (-6.5, -3.25) {$\alpha_2$};
		\node [style=none] (33) at (-7, -4.25) {...};
		\node [style=none] (34) at (-7.5, -4.25) {};
		\node [style=Z phase dot] (35) at (-3.25, -3.25) {$\ \alpha_{n\!-\!1}\ $};
		\node [style=none] (36) at (-3.25, -4.25) {};
		\node [style=none] (37) at (-2.25, -4.25) {};
		\node [style=none] (38) at (-2.75, -4.25) {...};
		\node [style=Z phase dot] (39) at (9.5, -2.75) {$\ \alpha_2\!\mp\!\frac\pi2\ $};
		\node [style=none] (40) at (9, -3.75) {...};
		\node [style=none] (41) at (8.5, -3.75) {};
		\node [style=none] (42) at (9.5, -3.75) {};
		\node [style=Z phase dot] (43) at (12.5, -2.75) {$\ \alpha_{n\!-\!1} \!\mp\! \frac\pi2\ $};
		\node [style=none] (44) at (13.5, -3.75) {};
		\node [style=none] (45) at (13, -3.75) {...};
		\node [style=none] (46) at (12.5, -3.75) {};
		\node [style=none] (47) at (11, -1) {...};
		\node [style=none] (48) at (3.5, -1.5) {$e^{\pm i \pi/4} \sqrt{2}^{\frac{(n-1)(n-2)}{2}}$};
	\end{pgfonlayer}
	\begin{pgfonlayer}{edgelayer}
		\draw (5.center) to (2);
		\draw (6.center) to (2);
		\draw (7.center) to (4);
		\draw (8.center) to (4);
		\draw [style=hadamard edge] (2) to (0);
		\draw [style=hadamard edge] (4) to (0);
		\draw (18.center) to (24);
		\draw (26.center) to (24);
		\draw (20.center) to (30);
		\draw (17.center) to (30);
		\draw (34.center) to (32);
		\draw (31.center) to (32);
		\draw (36.center) to (35);
		\draw (37.center) to (35);
		\draw (41.center) to (39);
		\draw (42.center) to (39);
		\draw (46.center) to (43);
		\draw (44.center) to (43);
		\draw [style=hadamard edge] (24) to (30);
		\draw [style=hadamard edge] (30) to (43);
		\draw [style=hadamard edge] (43) to (39);
		\draw [style=hadamard edge] (39) to (24);
		\draw [style=hadamard edge] (24) to (43);
		\draw [style=hadamard edge] (39) to (30);
		\draw [style=hadamard edge] (0) to (32);
		\draw [style=hadamard edge] (0) to (35);
	\end{pgfonlayer}
\end{tikzpicture}
\end{equation}
While this rule might introduce a lot of new edges between vertices, the important point is that the resulting diagram has one less vertex. Hence, this can only be applied a finite number of times before it terminates.

The second rewrite rule removes a pair of connected internal spiders that both have a phase of $0$ or $\pi$, again at the cost of introducing some new edges:
\begin{equation}\label{eq:pivot-simp}
  \begin{tikzpicture}
	\begin{pgfonlayer}{nodelayer}
		\node [style=Z phase dot] (0) at (-8.5, 2.25) {$j\pi$};
		\node [style=Z phase dot] (2) at (-10.25, 1.75) {$\alpha_1$};
		\node [style=none] (16) at (0, 0) {$=$};
		\node [style=Z phase dot] (32) at (-10.25, 0.25) {$\alpha_n$};
		\node [style=Z phase dot] (37) at (-6.5, -1.25) {$\beta_m$};
		\node [style=Z phase dot] (41) at (-6.5, 0.25) {$\beta_1$};
		\node [style=Z phase dot] (45) at (-3, 1.75) {$\gamma_1$};
		\node [style=Z phase dot] (49) at (-3, 0.25) {$\gamma_l$};
		\node [style=Z phase dot] (51) at (-4.5, 2.25) {$k\pi$};
		\node [style=none] (52) at (-10.25, 1) {...};
		\node [style=none] (53) at (-6.5, -0.5) {...};
		\node [style=none] (54) at (-3, 1) {...};
		\node [style=Z phase dot] (55) at (7.75, 0.75) {$\ \alpha_n+k\pi\ $};
		\node [style=Z phase dot] (56) at (11.5, -1.25) {$\ \beta_1+(j+k+1)\pi\ $};
		\node [style=none] (57) at (11.5, -2) {...};
		\node [style=Z phase dot] (58) at (11.5, -2.75) {$\ \beta_m+(j+k+1)\pi\ $};
		\node [style=Z phase dot] (59) at (15.25, 2.25) {$\ \gamma_1+j\pi\ $};
		\node [style=Z phase dot] (60) at (7.75, 2.25) {$\ \alpha_1+k\pi\ $};
		\node [style=none] (61) at (15.25, 1.5) {...};
		\node [style=none] (62) at (7.75, 1.5) {...};
		\node [style=Z phase dot] (63) at (15.25, 0.75) {$\ \gamma_l+j\pi\ $};
		\node [style=none] (64) at (-9.75, 2.5) {};
		\node [style=none] (65) at (-10.75, 2.5) {};
		\node [style=none] (66) at (-10.25, 2.5) {...};
		\node [style=none] (67) at (-10.75, -0.5) {};
		\node [style=none] (68) at (-10.25, -0.5) {...};
		\node [style=none] (69) at (-9.75, -0.5) {};
		\node [style=none] (70) at (-7, -2) {};
		\node [style=none] (71) at (-6.5, -2) {...};
		\node [style=none] (72) at (-6, -2) {};
		\node [style=none] (73) at (-7.25, 1.25) {};
		\node [style=none] (74) at (-6.5, 1) {...};
		\node [style=none] (75) at (-5.75, 1.25) {};
		\node [style=none] (76) at (-3.5, 2.5) {};
		\node [style=none] (77) at (-3, 2.5) {...};
		\node [style=none] (78) at (-2.5, 2.5) {};
		\node [style=none] (79) at (-3, -0.5) {...};
		\node [style=none] (80) at (-3.5, -0.5) {};
		\node [style=none] (81) at (-2.5, -0.5) {};
		\node [style=none] (82) at (7.25, 3) {};
		\node [style=none] (83) at (7.75, 3) {...};
		\node [style=none] (84) at (8.25, 3) {};
		\node [style=none] (85) at (7.75, 0) {...};
		\node [style=none] (86) at (7.25, 0) {};
		\node [style=none] (87) at (8.25, 0) {};
		\node [style=none] (88) at (11, -0.5) {};
		\node [style=none] (89) at (11.5, -0.5) {...};
		\node [style=none] (90) at (12, -0.5) {};
		\node [style=none] (91) at (11.5, -3.5) {...};
		\node [style=none] (92) at (11, -3.5) {};
		\node [style=none] (93) at (12, -3.5) {};
		\node [style=none] (94) at (14.75, 3) {};
		\node [style=none] (95) at (15.25, 3) {...};
		\node [style=none] (96) at (15.75, 3) {};
		\node [style=none] (97) at (15.25, 0) {...};
		\node [style=none] (98) at (14.75, 0) {};
		\node [style=none] (99) at (15.75, 0) {};
		\node [style=none] (112) at (3.5, 0) {$(-1)^{jk} \sqrt{2}^{E}$};
	\end{pgfonlayer}
	\begin{pgfonlayer}{edgelayer}
		\draw [style=hadamard edge] (2) to (0);
		\draw [style=hadamard edge] (0) to (32);
		\draw [style=hadamard edge] (0) to (51);
		\draw [style=hadamard edge, bend right] (0) to (37);
		\draw [style=hadamard edge, bend right] (0) to (41);
		\draw [style=hadamard edge, bend left] (51) to (37);
		\draw [style=hadamard edge, bend left, looseness=0.75] (51) to (41);
		\draw [style=hadamard edge] (51) to (45);
		\draw [style=hadamard edge] (51) to (49);
		\draw [style=hadamard edge] (60) to (56);
		\draw [style=hadamard edge] (60) to (58);
		\draw [style=hadamard edge] (55) to (56);
		\draw [style=hadamard edge] (55) to (58);
		\draw [style=hadamard edge] (59) to (56);
		\draw [style=hadamard edge] (59) to (58);
		\draw [style=hadamard edge] (56) to (63);
		\draw [style=hadamard edge] (63) to (58);
		\draw [style=hadamard edge] (60) to (59);
		\draw [style=hadamard edge] (60) to (63);
		\draw [style=hadamard edge] (55) to (63);
		\draw [style=hadamard edge] (55) to (59);
		\draw (64.center) to (2);
		\draw (2) to (65.center);
		\draw (67.center) to (32);
		\draw (69.center) to (32);
		\draw (70.center) to (37);
		\draw (72.center) to (37);
		\draw (41) to (75.center);
		\draw (41) to (73.center);
		\draw (45) to (78.center);
		\draw (45) to (76.center);
		\draw (49) to (80.center);
		\draw (49) to (81.center);
		\draw (60) to (82.center);
		\draw (60) to (84.center);
		\draw (59) to (94.center);
		\draw (59) to (96.center);
		\draw (98.center) to (63);
		\draw (99.center) to (63);
		\draw (86.center) to (55);
		\draw (87.center) to (55);
		\draw (56) to (88.center);
		\draw (56) to (90.center);
		\draw (92.center) to (58);
		\draw (93.center) to (58);
	\end{pgfonlayer}
\end{tikzpicture}
\end{equation}
where $E = (n-1)m + (l-1)m + (n-1)(l-1)$.

These rules suffice to remove almost all internal \emph{Clifford} spiders (those whose phase is a multiple of $\frac\pi2$). Indeed, if we are dealing with a scalar diagram (for instance if we are calculating an amplitude), then the only Clifford spiders left after applying these rules wherever they can be applied are those spiders with a $0$ or $\pi$ phase that are connected only to \emph{non-Clifford} spiders.
Such spiders can also be removed, in a sense, by transforming them into \emph{phase gadgets} using a variation on the pivoting rule~\cite{kissinger2019tcount}:
\begin{equation}\label{eq:pivot-simp-gadget}
  \begin{tikzpicture}
	\begin{pgfonlayer}{nodelayer}
		\node [style=Z phase dot] (109) at (-7, 2.25) {$j\pi$};
		\node [style=Z phase dot] (110) at (-8.5, 1.75) {$\alpha_1$};
		\node [style=none] (111) at (0, 0) {$=$};
		\node [style=Z phase dot] (112) at (-8.5, 0.25) {$\alpha_n$};
		\node [style=Z phase dot] (113) at (-6.5, -1.25) {$\beta_1$};
		\node [style=Z phase dot] (114) at (-4.25, -1.25) {$\beta_m$};
		\node [style=Z phase dot] (115) at (-2.25, 1.75) {$\gamma_1$};
		\node [style=Z phase dot] (116) at (-2.25, 0.25) {$\gamma_l$};
		\node [style=Z phase dot] (117) at (-3.5, 2.25) {$\alpha$};
		\node [style=none] (118) at (-5.375, -1.25) {...};
		\node [style=Z phase dot] (119) at (6.75, 0.75) {$\ \alpha_n+k\pi\ $};
		\node [style=Z phase dot] (120) at (7, -2.25) {$\ \beta_1+(j+k+1)\pi\ $};
		\node [style=none] (121) at (9.75, -2.25) {...};
		\node [style=Z phase dot] (122) at (12.5, -2.25) {$\ \beta_m+(j+k+1)\pi\ $};
		\node [style=Z phase dot] (123) at (13.25, 2.25) {$\ \gamma_1+j\pi\ $};
		\node [style=Z phase dot] (124) at (6.75, 2.25) {$\ \alpha_1+k\pi\ $};
		\node [style=Z phase dot] (125) at (13.25, 0.75) {$\ \gamma_l+j\pi\ $};
		\node [style=none] (126) at (-9.5, 2.375) {};
		\node [style=none] (127) at (-9.5, 1.25) {};
		\node [style=none] (128) at (-9.5, 0.75) {};
		\node [style=none] (129) at (-9.5, -0.5) {};
		\node [style=none] (130) at (-7, -2.25) {};
		\node [style=none] (131) at (-6.5, -2) {...};
		\node [style=none] (132) at (-6, -2.25) {};
		\node [style=none] (133) at (-4.75, -2.25) {};
		\node [style=none] (134) at (-4.25, -2) {...};
		\node [style=none] (135) at (-3.75, -2.25) {};
		\node [style=none] (136) at (-1.25, 2.25) {};
		\node [style=none] (137) at (-1.25, 1) {};
		\node [style=none] (138) at (-1.25, 0.75) {};
		\node [style=none] (139) at (-1.25, -0.5) {};
		\node [style=none] (140) at (5.25, 1.75) {};
		\node [style=none] (141) at (5.25, 3) {};
		\node [style=none] (142) at (5.25, 1.25) {};
		\node [style=none] (143) at (5.25, 0) {};
		\node [style=none] (144) at (6.5, -3.25) {};
		\node [style=none] (145) at (7, -3) {...};
		\node [style=none] (146) at (7.5, -3.25) {};
		\node [style=none] (147) at (12.5, -3) {...};
		\node [style=none] (148) at (12, -3.25) {};
		\node [style=none] (149) at (13, -3.25) {};
		\node [style=none] (150) at (14.75, 3) {};
		\node [style=none] (151) at (14.75, 1.5) {};
		\node [style=none] (152) at (14.75, -0.25) {};
		\node [style=none] (153) at (14.75, 1.25) {};
		\node [style=none, rotate=90] (166) at (14.625, 0.625) {...};
		\node [style=none, rotate=90] (167) at (14.625, 2.375) {...};
		\node [style=none, rotate=90] (168) at (13.25, 1.5) {...};
		\node [style=none, rotate=90] (169) at (6.75, 1.5) {...};
		\node [style=none, rotate=90] (170) at (5.375, 0.625) {...};
		\node [style=none, rotate=90] (171) at (5.375, 2.375) {...};
		\node [style=none, rotate=90] (172) at (-9.375, 1.8) {...};
		\node [style=none, rotate=90] (173) at (-9.375, 0.125) {...};
		\node [style=none, rotate=90] (174) at (-1.5, 1.7) {...};
		\node [style=none, rotate=90] (175) at (-1.5, 0.2) {...};
		\node [style=none, rotate=90] (176) at (-8.5, 1.025) {...};
		\node [style=none, rotate=90] (177) at (-2.25, 1.025) {...};
		\node [style=Z phase dot] (178) at (11.5, 4.5) {\ (-1)$^j\alpha$\ };
		\node [style=Z dot] (179) at (11.5, 3.5) {};
		\node [style=none] (181) at (2.5, 0) {$e^{ij\alpha}\sqrt{2}^{E'}$};
	\end{pgfonlayer}
	\begin{pgfonlayer}{edgelayer}
		\draw [style=hadamard edge] (110) to (109);
		\draw [style=hadamard edge] (109) to (112);
		\draw [style=hadamard edge] (109) to (117);
		\draw [style=hadamard edge] (109) to (113);
		\draw [style=hadamard edge] (109) to (114);
		\draw [style=hadamard edge] (117) to (113);
		\draw [style=hadamard edge] (117) to (114);
		\draw [style=hadamard edge] (117) to (115);
		\draw [style=hadamard edge] (117) to (116);
		\draw [style=hadamard edge, bend left=60, looseness=1.25] (124) to (120);
		\draw [style=hadamard edge] (124) to (122);
		\draw [style=hadamard edge] (119) to (120);
		\draw [style=hadamard edge] (119) to (122);
		\draw [style=hadamard edge] (123) to (120);
		\draw [style=hadamard edge, bend right=60, looseness=1.25] (123) to (122);
		\draw [style=hadamard edge] (120) to (125);
		\draw [style=hadamard edge] (125) to (122);
		\draw [style=hadamard edge] (124) to (123);
		\draw [style=hadamard edge] (124) to (125);
		\draw [style=hadamard edge] (119) to (125);
		\draw [style=hadamard edge] (119) to (123);
		\draw (126.center) to (110);
		\draw (110) to (127.center);
		\draw (128.center) to (112);
		\draw (129.center) to (112);
		\draw (130.center) to (113);
		\draw (132.center) to (113);
		\draw (114) to (135.center);
		\draw (114) to (133.center);
		\draw (115) to (137.center);
		\draw (115) to (136.center);
		\draw (116) to (138.center);
		\draw (116) to (139.center);
		\draw (124) to (140.center);
		\draw (124) to (141.center);
		\draw (123) to (150.center);
		\draw (123) to (151.center);
		\draw (152.center) to (125);
		\draw (153.center) to (125);
		\draw (142.center) to (119);
		\draw (143.center) to (119);
		\draw (120) to (144.center);
		\draw (120) to (146.center);
		\draw (148.center) to (122);
		\draw (149.center) to (122);
		\draw [style=hadamard edge] (179) to (178);
		\draw [style=hadamard edge] (179) to (124);
		\draw [style=hadamard edge] (179) to (119);
		\draw [style=hadamard edge] (179) to (120);
		\draw [style=hadamard edge] (179) to (122);
	\end{pgfonlayer}
\end{tikzpicture}
\end{equation}
where $E' = E + m + n - 1 = (n-1)m + lm + (n-1)l$.

A phase gadget is simply a phaseless spider to which is attached a 1-ary spider:
\begin{equation}
  \begin{tikzpicture}
	\begin{pgfonlayer}{nodelayer}
		\node [style=Z dot] (5) at (0.25, 0) {};
		\node [style=none] (6) at (1.25, -0.75) {};
		\node [style=none] (7) at (1.25, 0.75) {};
		\node [style=none, rotate=90] (8) at (1, 0) {...};
		\node [style=Z phase dot] (10) at (-1.25, 0) {$\alpha$};
	\end{pgfonlayer}
	\begin{pgfonlayer}{edgelayer}
		\draw (6.center) to (5);
		\draw (5) to (7.center);
		\draw [style=hadamard edge] (5) to (10);
	\end{pgfonlayer}
\end{tikzpicture}
\end{equation}
For many purposes it is helpful to consider the two spiders of a phase gadget together as a single component. In particular, for the application of a pivot~\eqref{eq:pivot-simp} we do not consider the `base' (i.e. the phase-free spider) in a phase gadget a valid target for pivoting, because doing so could result in a loop of applications of~\eqref{eq:pivot-simp} and~\eqref{eq:pivot-simp-gadget}.

So far we have only shown how we can remove all the Clifford spiders by either removing them directly or making them part of a phase gadget. We now show two rewrite rules that apply to phase gadgets that allow us to merge or kill some non-Clifford phases in the diagram.
The first allows us to fuse a phase gadget whose base has only a single neighbour into this neighbour.
\begin{equation}
  \begin{tikzpicture}
	\begin{pgfonlayer}{nodelayer}
		\node [style=Z phase dot] (0) at (-6, 0) {$\alpha$};
		\node [style=Z phase dot] (1) at (-3, 0) {$\beta$};
		\node [style=Z dot] (2) at (-4.5, 0) {};
		\node [style=none] (3) at (-1.5, 1) {};
		\node [style=none] (4) at (-1.5, -1) {};
		\node [style=none, rotate=90] (8) at (-2, 0) {...};
		\node [style=none] (18) at (0, 0) {=};
		\node [style=Z phase dot] (19) at (3, 0) {$\ \alpha+\beta\ $};
		\node [style=none] (21) at (4.75, 1) {};
		\node [style=none] (22) at (4.75, -1) {};
		\node [style=none, rotate=90] (26) at (4.5, 0) {...};
	\end{pgfonlayer}
	\begin{pgfonlayer}{edgelayer}
		\draw (1) to (3.center);
		\draw (1) to (4.center);
		\draw [style=hadamard edge] (0) to (2);
		\draw [style=hadamard edge] (2) to (1);
		\draw (19) to (21.center);
		\draw (19) to (22.center);
	\end{pgfonlayer}
\end{tikzpicture}
\end{equation}
The second allows us to merge phase gadgets whose bases have exactly the same neighbourhood:
\begin{equation}
  \begin{tikzpicture}
	\begin{pgfonlayer}{nodelayer}
		\node [style=Z dot] (51) at (-4.25, -0.75) {};
		\node [style=Z phase dot] (52) at (-5.25, -0.75) {$\alpha$};
		\node [style=none] (53) at (-3, 1.75) {};
		\node [style=none] (55) at (-1.5, 1.75) {};
		\node [style=Z dot] (57) at (-2.25, 1) {};
		\node [style=none] (59) at (-3, -2) {};
		\node [style=none] (60) at (-1.5, -2) {};
		\node [style=Z dot] (61) at (-2.25, -1.25) {};
		\node [style=Z dot] (62) at (-4.25, 0.75) {};
		\node [style=Z phase dot] (63) at (-5.25, 0.75) {$\beta$};
		\node [style=Z phase dot] (64) at (-2.25, 1) {$\alpha_1$};
		\node [style=Z phase dot] (66) at (-2.25, -1.25) {$\alpha_n$};
		\node [style=none, rotate=90] (131) at (-2.25, 0) {...};
		\node [style=Z dot] (132) at (7.25, 0) {};
		\node [style=Z phase dot] (133) at (5.5, 0) {$\ \alpha+\beta\ $};
		\node [style=none] (134) at (9, 1.75) {};
		\node [style=none] (135) at (7.5, 1.75) {};
		\node [style=Z phase dot] (136) at (8.25, 1) {$\alpha_1$};
		\node [style=none] (137) at (7.5, -1.75) {};
		\node [style=none] (138) at (9, -1.75) {};
		\node [style=Z dot] (139) at (8.25, -1) {};
		\node [style=Z phase dot] (143) at (8.25, -1) {$\alpha_n$};
		\node [style=none, rotate=90] (144) at (8.25, 0) {...};
		\node [style=none] (145) at (0, 0) {=};
		\node [style=none] (146) at (-2.25, -2) {...};
		\node [style=none] (147) at (-2.25, 1.75) {...};
		\node [style=none] (148) at (8.25, 1.75) {...};
		\node [style=none] (149) at (8.25, -1.75) {...};
		\node [style=none] (150) at (2.5, 0) {$\left(\frac{1}{\sqrt{2}}\right)^{n-1}$};
	\end{pgfonlayer}
	\begin{pgfonlayer}{edgelayer}
		\draw [style=hadamard edge] (51) to (52);
		\draw (53.center) to (57);
		\draw (57) to (55.center);
		\draw [style=hadamard edge] (51) to (57);
		\draw (59.center) to (61);
		\draw (61) to (60.center);
		\draw [style=hadamard edge] (61) to (51);
		\draw [style=hadamard edge] (62) to (63);
		\draw [style=hadamard edge] (62) to (64);
		\draw [style=hadamard edge] (66) to (62);
		\draw [style=hadamard edge] (132) to (133);
		\draw (134.center) to (136);
		\draw (136) to (135.center);
		\draw [style=hadamard edge] (132) to (136);
		\draw (137.center) to (139);
		\draw (139) to (138.center);
		\draw [style=hadamard edge] (139) to (132);
	\end{pgfonlayer}
\end{tikzpicture}
\end{equation}

Note that these last rules can combine $\frac\pi4$ phases together so that they become multiples of $\frac\pi2$, meaning that they become targets for the application of~\eqref{eq:lc-simp} and \eqref{eq:pivot-simp}. We hence repeat this procedure until none of the rewrite rules match. In the case we are dealing with a scalar diagram we will then have a diagram where every spider has a non-Clifford phase or is part of a phase gadget whose phase is non-Clifford. If we started with a circuit with $t$ non-Clifford gates, we will hence get a diagram of size $O(t)$. Note that if the original diagram had $N$ spiders, that this procedure takes at most $O(N^3)$ elementary graph operations (vertex deletions, edge toggles), as each rewrite can affect at most $N^2$ edges, and we apply at most $N$ of these rewrite rules as each deletes a vertex.


\section{Methods}\label{sec:method}

Our method for classically simulating a quantum circuit consists of essentially combining the work done on stabiliser decompositions with that done on ZX-diagram simplification. We first describe the procedure for calculating a single amplitude then proceed to the calculation of marginal probabilities.

Let $U$ be a unitary given as an $n$-qubit Clifford+T circuit that consists of $N$ gates of which $t$ are T gates.
Our goal is to evaluate the amplitude $\bra{x_1\cdots x_n} U \ket{0\cdots 0}$ for some fixed computational basis effect $\bra{x_1\cdots x_n}$.

The first step is to write this amplitude as a ZX-diagram. This is straightforward. We simply translate each of the gates in $U$ to spiders and compose the resulting circuit-like ZX-diagram with the necessary basis states and effects $\ket{0}$, $\ket{1}$, $\bra{0}$, and $\bra{1}$ (which can also be expressed as spiders, as described in Example~\ref{ex:basic-maps-and-states}). This results in the following ZX-diagram:
\begin{equation}
  \begin{tikzpicture}
	\begin{pgfonlayer}{nodelayer}
		\node [style=X dot] (32) at (-6.5, 2.75) {};
		\node [style=X phase dot] (33) at (1.75, 2.75) {$\ x_1\pi~$};
		\node [style=X dot] (34) at (-6.5, 1.25) {};
		\node [style=X phase dot] (35) at (1.75, 1.25) {$\ x_2\pi~$};
		\node [style=X dot] (36) at (-6.5, -0.25) {};
		\node [style=X phase dot] (37) at (1.75, -0.25) {$\ x_3\pi~$};
		\node [style=X dot] (53) at (-6.5, -2.25) {};
		\node [style=X phase dot] (54) at (1.75, -2.25) {$\ x_n\pi~$};
		\node [style=none] (55) at (-4.5, 3.25) {};
		\node [style=none] (56) at (-4.5, -2.75) {};
		\node [style=none] (57) at (-0.5, -2.75) {};
		\node [style=none] (58) at (-0.5, 3.25) {};
		\node [style=none] (59) at (-4.5, 2.75) {};
		\node [style=none] (60) at (-4.5, 1.25) {};
		\node [style=none] (61) at (-4.5, -0.25) {};
		\node [style=none] (63) at (-4.5, -2.25) {};
		\node [style=none] (64) at (-0.5, -2.25) {};
		\node [style=none] (66) at (-0.5, -0.25) {};
		\node [style=none] (67) at (-0.5, 1.25) {};
		\node [style=none] (68) at (-0.5, 2.75) {};
		\node [style=none] (69) at (-2.5, 0.375) {$U$};
		\node [style=none, rotate=90] (168) at (-5.25, -1.25) {$\cdots$};
		\node [style=none, rotate=90] (169) at (0.75, -1.25) {$\cdots$};
		\node [style=none] (170) at (-8.75, 0) {$\left(\frac{1}{\sqrt{2}}\right)^{2n}$};
	\end{pgfonlayer}
	\begin{pgfonlayer}{edgelayer}
		\draw (55.center) to (58.center);
		\draw (58.center) to (57.center);
		\draw (57.center) to (56.center);
		\draw (56.center) to (55.center);
		\draw (68.center) to (33);
		\draw (53) to (63.center);
		\draw (36) to (61.center);
		\draw (60.center) to (34);
		\draw (32) to (59.center);
		\draw (64.center) to (54);
		\draw (37) to (66.center);
		\draw (67.center) to (35);
	\end{pgfonlayer}
\end{tikzpicture}
\end{equation}
Since this diagram has no inputs or outputs, it represents a $2^0 \times 2^0$ matrix, i.e. a scalar.

We then simplify this diagram using the procedure described in Section~\ref{sec:simplify-ZX}. This takes at most $O(N^3)$ elementary graph operations and results in a diagram with $O(t)$ spiders, where each T gate that remains after the simplification contributes either a single spider or a phase gadget. For example, after simplification, a circuit could yield a reduced ZX-diagram that looks like this:
\begin{equation}
  \begin{tikzpicture}
	\begin{pgfonlayer}{nodelayer}
		\node [style=Z phase dot] (41) at (-11.25, -2.25) {$\frac{3\pi}{4}$};
		\node [style=Z dot] (45) at (-12, 2.5) {};
		\node [style=Z dot] (117) at (-10.25, 2.5) {};
		\node [style=Z phase dot] (119) at (-6.25, 0.25) {$\frac{3\pi}{4}$};
		\node [style=Z dot] (123) at (-8.75, 2.5) {};
		\node [style=Z dot] (127) at (-7.75, 2.5) {};
		\node [style=Z phase dot] (143) at (6, -0.75) {$\frac{\pi}{4}$};
		\node [style=Z dot] (147) at (-6.25, 2.5) {};
		\node [style=Z dot] (155) at (-4.75, 2.5) {};
		\node [style=Z dot] (165) at (-3.25, 2.5) {};
		\node [style=Z phase dot] (178) at (8.5, -0.5) {$\frac{\pi}{4}$};
		\node [style=Z dot] (183) at (-1.5, 2.5) {};
		\node [style=Z dot] (186) at (0.5, 2.5) {};
		\node [style=Z phase dot] (188) at (0.75, -1) {$\frac{7\pi}{4}$};
		\node [style=Z dot] (189) at (2.5, 2.5) {};
		\node [style=Z dot] (193) at (3.5, 2.5) {};
		\node [style=Z phase dot] (236) at (14, -3.25) {$\frac{5\pi}{4}$};
		\node [style=Z dot] (245) at (5.25, 2.5) {};
		\node [style=Z dot] (250) at (6.25, 2.5) {};
		\node [style=Z phase dot] (253) at (3.75, -3.5) {$\frac{5\pi}{4}$};
		\node [style=Z dot] (261) at (7.75, 2.5) {};
		\node [style=Z dot] (272) at (9.25, 2.5) {};
		\node [style=Z phase dot] (285) at (16.5, -1.25) {$\frac{7\pi}{4}$};
		\node [style=Z dot] (289) at (11, 2.5) {};
		\node [style=Z dot] (293) at (12, 2.5) {};
		\node [style=Z dot] (328) at (13.5, 2.5) {};
		\node [style=Z dot] (331) at (15.5, 2.5) {};
		\node [style=Z dot] (334) at (17.5, 2.5) {};
		\node [style=Z phase dot] (337) at (15.25, -5) {$\frac{\pi}{4}$};
		\node [style=Z phase dot] (391) at (11, 3.5) {$\frac{\pi}{4}$};
		\node [style=Z phase dot] (392) at (-6.25, 3.5) {$\frac{7\pi}{4}$};
		\node [style=Z phase dot] (397) at (3.5, 3.5) {$\frac{\pi}{4}$};
		\node [style=Z phase dot] (402) at (11.5, -0.5) {$\frac{5\pi}{4}$};
		\node [style=Z phase dot] (407) at (-4.75, 3.5) {$\frac{\pi}{4}$};
		\node [style=Z phase dot] (408) at (0.5, 3.5) {$\frac{7\pi}{4}$};
		\node [style=Z phase dot] (411) at (7.75, 3.5) {$\frac{\pi}{4}$};
		\node [style=Z phase dot] (413) at (13.5, 3.5) {$\frac{\pi}{4}$};
		\node [style=Z phase dot] (416) at (17.5, 3.5) {$\frac{7\pi}{4}$};
		\node [style=Z phase dot] (421) at (2.5, 3.5) {$\frac{\pi}{4}$};
		\node [style=Z phase dot] (424) at (15.5, 3.5) {$\frac{3\pi}{4}$};
		\node [style=Z phase dot] (425) at (-3.25, 3.5) {$\frac{5\pi}{4}$};
		\node [style=Z phase dot] (426) at (5.25, 3.5) {$\frac{7\pi}{4}$};
		\node [style=Z phase dot] (430) at (9.25, 3.5) {$\frac{7\pi}{4}$};
		\node [style=Z phase dot] (434) at (-1.5, 3.5) {$\frac{\pi}{4}$};
		\node [style=Z phase dot] (435) at (-8.75, 3.5) {$\frac{\pi}{4}$};
		\node [style=Z phase dot] (436) at (6.25, 3.5) {$\frac{\pi}{4}$};
		\node [style=Z phase dot] (438) at (-6.5, -3.75) {$\frac{3\pi}{4}$};
		\node [style=Z phase dot] (439) at (-10.25, 3.5) {$\frac{7\pi}{4}$};
		\node [style=Z phase dot] (440) at (-12, 3.5) {$\frac{5\pi}{4}$};
		\node [style=Z phase dot] (443) at (12, 3.5) {$\frac{5\pi}{4}$};
		\node [style=Z phase dot] (447) at (-3, -1.5) {$\frac{7\pi}{4}$};
		\node [style=Z phase dot] (454) at (10, -3.75) {$\frac{\pi}{4}$};
		\node [style=Z phase dot] (461) at (-7.75, 3.5) {$\frac{5\pi}{4}$};
	\end{pgfonlayer}
	\begin{pgfonlayer}{edgelayer}
		\draw [style=hadamard edge] (41) to (454);
		\draw [style=hadamard edge] (41) to (447);
		\draw [style=hadamard edge] (41) to (45);
		\draw [style=hadamard edge] (41) to (402);
		\draw [style=hadamard edge] (41) to (117);
		\draw [style=hadamard edge] (41) to (178);
		\draw [style=hadamard edge] (41) to (438);
		\draw [style=hadamard edge] (41) to (188);
		\draw [style=hadamard edge] (41) to (253);
		\draw [style=hadamard edge] (41) to (127);
		\draw [style=hadamard edge] (45) to (438);
		\draw [style=hadamard edge] (45) to (440);
		\draw [style=hadamard edge] (117) to (119);
		\draw [style=hadamard edge] (117) to (438);
		\draw [style=hadamard edge] (117) to (439);
		\draw [style=hadamard edge] (119) to (143);
		\draw [style=hadamard edge] (119) to (337);
		\draw [style=hadamard edge] (119) to (402);
		\draw [style=hadamard edge] (119) to (123);
		\draw [style=hadamard edge] (119) to (447);
		\draw [style=hadamard edge] (119) to (438);
		\draw [style=hadamard edge] (119) to (127);
		\draw [style=hadamard edge] (119) to (253);
		\draw [style=hadamard edge] (123) to (438);
		\draw [style=hadamard edge] (123) to (435);
		\draw [style=hadamard edge] (127) to (461);
		\draw [style=hadamard edge] (143) to (193);
		\draw [style=hadamard edge] (143) to (183);
		\draw [style=hadamard edge] (143) to (337);
		\draw [style=hadamard edge] (143) to (155);
		\draw [style=hadamard edge] (143) to (165);
		\draw [style=hadamard edge] (143) to (438);
		\draw [style=hadamard edge] (143) to (186);
		\draw [style=hadamard edge] (147) to (454);
		\draw [style=hadamard edge] (147) to (392);
		\draw [style=hadamard edge] (147) to (402);
		\draw [style=hadamard edge] (147) to (253);
		\draw [style=hadamard edge] (147) to (447);
		\draw [style=hadamard edge] (155) to (454);
		\draw [style=hadamard edge] (155) to (402);
		\draw [style=hadamard edge] (155) to (407);
		\draw [style=hadamard edge] (155) to (253);
		\draw [style=hadamard edge] (155) to (447);
		\draw [style=hadamard edge] (165) to (454);
		\draw [style=hadamard edge] (165) to (402);
		\draw [style=hadamard edge] (165) to (425);
		\draw [style=hadamard edge] (165) to (253);
		\draw [style=hadamard edge] (178) to (193);
		\draw [style=hadamard edge] (178) to (454);
		\draw [style=hadamard edge] (178) to (236);
		\draw [style=hadamard edge] (178) to (253);
		\draw [style=hadamard edge] (178) to (438);
		\draw [style=hadamard edge] (178) to (183);
		\draw [style=hadamard edge] (178) to (189);
		\draw [style=hadamard edge] (183) to (434);
		\draw [style=hadamard edge] (183) to (188);
		\draw [style=hadamard edge] (183) to (447);
		\draw [style=hadamard edge] (186) to (188);
		\draw [style=hadamard edge] (186) to (408);
		\draw [style=hadamard edge] (186) to (447);
		\draw [style=hadamard edge] (188) to (454);
		\draw [style=hadamard edge] (188) to (236);
		\draw [style=hadamard edge] (188) to (253);
		\draw [style=hadamard edge] (188) to (438);
		\draw [style=hadamard edge] (188) to (189);
		\draw [style=hadamard edge] (189) to (421);
		\draw [style=hadamard edge] (193) to (397);
		\draw [style=hadamard edge] (193) to (447);
		\draw [style=hadamard edge] (236) to (261);
		\draw [style=hadamard edge] (236) to (454);
		\draw [style=hadamard edge] (236) to (328);
		\draw [style=hadamard edge] (236) to (334);
		\draw [style=hadamard edge] (236) to (402);
		\draw [style=hadamard edge] (236) to (285);
		\draw [style=hadamard edge] (236) to (293);
		\draw [style=hadamard edge] (236) to (289);
		\draw [style=hadamard edge] (236) to (245);
		\draw [style=hadamard edge] (236) to (250);
		\draw [style=hadamard edge] (236) to (447);
		\draw [style=hadamard edge] (245) to (402);
		\draw [style=hadamard edge] (245) to (426);
		\draw [style=hadamard edge] (245) to (253);
		\draw [style=hadamard edge] (250) to (454);
		\draw [style=hadamard edge] (250) to (436);
		\draw [style=hadamard edge] (250) to (253);
		\draw [style=hadamard edge] (253) to (272);
		\draw [style=hadamard edge] (253) to (402);
		\draw [style=hadamard edge] (253) to (285);
		\draw [style=hadamard edge] (253) to (293);
		\draw [style=hadamard edge] (253) to (438);
		\draw [style=hadamard edge] (253) to (447);
		\draw [style=hadamard edge] (253) to (454);
		\draw [style=hadamard edge] (253) to (331);
		\draw [style=hadamard edge] (253) to (334);
		\draw [style=hadamard edge] (253) to (337);
		\draw [style=hadamard edge] (261) to (454);
		\draw [style=hadamard edge] (261) to (402);
		\draw [style=hadamard edge] (261) to (411);
		\draw [style=hadamard edge] (272) to (285);
		\draw [style=hadamard edge] (272) to (430);
		\draw [style=hadamard edge] (285) to (454);
		\draw [style=hadamard edge] (285) to (289);
		\draw [style=hadamard edge] (285) to (293);
		\draw [style=hadamard edge] (289) to (391);
		\draw [style=hadamard edge] (289) to (402);
		\draw [style=hadamard edge] (293) to (402);
		\draw [style=hadamard edge] (293) to (443);
		\draw [style=hadamard edge] (328) to (413);
		\draw [style=hadamard edge] (328) to (337);
		\draw [style=hadamard edge] (328) to (402);
		\draw [style=hadamard edge] (331) to (424);
		\draw [style=hadamard edge] (331) to (337);
		\draw [style=hadamard edge] (334) to (416);
		\draw [style=hadamard edge] (334) to (337);
		\draw [style=hadamard edge] (334) to (402);
		\draw [style=hadamard edge] (337) to (438);
		\draw [style=hadamard edge] (337) to (447);
		\draw [style=hadamard edge] (402) to (438);
		\draw [style=hadamard edge] (438) to (447);
		\draw [style=hadamard edge] (447) to (454);
	\end{pgfonlayer}
\end{tikzpicture}
\end{equation}

Now comes the step where the magic happens. In~\cite{bravyi2016trading}, equation~(11), Bravyi, Smith, and Smolin gave a decomposition of 6 $|T\>$ magic states into 7 stabiliser terms, which we will refer to as the \textit{BSS decomposition}. We can express a ZX-diagram version of the BSS decomposition as follows:
\begin{equation}\label{eq:magic-state-decomposition}
\begin{tikzpicture}
	\begin{pgfonlayer}{nodelayer}
		\node [style=none] (0) at (-12.25, 6.5) {$e^{i\pi/4}$};
		\node [style=Z phase dot] (1) at (-10.75, 6) {$\frac{\pi}{4}$};
		\node [style=none] (2) at (-10.75, 7.25) {};
		\node [style=Z phase dot] (3) at (-9.75, 6) {$\frac{\pi}{4}$};
		\node [style=none] (4) at (-9.75, 7.25) {};
		\node [style=Z phase dot] (5) at (-8.75, 6) {$\frac{\pi}{4}$};
		\node [style=none] (6) at (-8.75, 7.25) {};
		\node [style=Z phase dot] (7) at (-7.75, 6) {$\frac{\pi}{4}$};
		\node [style=none] (8) at (-7.75, 7.25) {};
		\node [style=Z phase dot] (9) at (-6.75, 6) {$\frac{\pi}{4}$};
		\node [style=none] (10) at (-6.75, 7.25) {};
		\node [style=Z phase dot] (11) at (-5.75, 6) {$\frac{\pi}{4}$};
		\node [style=none] (12) at (-5.75, 7.25) {};
		\node [style=none] (13) at (-3.75, 6.25) {=};
		\node [style=none] (14) at (-11, 3) {$-\frac{1+\sqrt{2}}{4}$};
		\node [style=Z dot] (15) at (-9.25, 2.5) {};
		\node [style=none] (16) at (-9.25, 4) {};
		\node [style=Z dot] (17) at (-8.25, 2.5) {};
		\node [style=none] (18) at (-8.25, 4) {};
		\node [style=Z dot] (19) at (-7.25, 2.5) {};
		\node [style=none] (20) at (-7.25, 4) {};
		\node [style=Z dot] (21) at (-6.25, 2.5) {};
		\node [style=none] (22) at (-6.25, 4) {};
		\node [style=Z dot] (23) at (-5.25, 2.5) {};
		\node [style=none] (24) at (-5.25, 4) {};
		\node [style=Z dot] (25) at (-4.25, 2.5) {};
		\node [style=none] (26) at (-4.25, 4) {};
		\node [style=none] (27) at (-1.5, 3) {$\frac{1-\sqrt{2}}{4}$};
		\node [style=Z phase dot] (28) at (0.25, 2.5) {$\pi$};
		\node [style=none] (29) at (0.25, 4) {};
		\node [style=Z phase dot] (30) at (1.25, 2.5) {$\pi$};
		\node [style=none] (31) at (1.25, 4) {};
		\node [style=Z phase dot] (32) at (2.25, 2.5) {$\pi$};
		\node [style=none] (33) at (2.25, 4) {};
		\node [style=Z phase dot] (34) at (3.25, 2.5) {$\pi$};
		\node [style=none] (35) at (3.25, 4) {};
		\node [style=Z phase dot] (36) at (4.25, 2.5) {$\pi$};
		\node [style=none] (37) at (4.25, 4) {};
		\node [style=Z phase dot] (38) at (5.25, 2.5) {$\pi$};
		\node [style=none] (39) at (5.25, 4) {};
		\node [style=none] (40) at (-3.25, 3) {+};
		\node [style=none] (41) at (-1.75, -0.25) {$-2i$};
		\node [style=Z phase dot] (42) at (2.75, -2.75) {$\pi$};
		\node [style=hadamard] (43) at (0.25, -1) {};
		\node [style=hadamard] (44) at (1.25, -1) {};
		\node [style=hadamard] (45) at (2.25, -1) {};
		\node [style=hadamard] (46) at (3.25, -1) {};
		\node [style=hadamard] (47) at (4.25, -1) {};
		\node [style=hadamard] (48) at (5.25, -1) {};
		\node [style=Z phase dot] (49) at (0.25, 0) {$\frac{\pi}{2}$};
		\node [style=Z phase dot] (50) at (1.25, 0) {$\frac{\pi}{2}$};
		\node [style=Z phase dot] (51) at (2.25, 0) {$\frac{\pi}{2}$};
		\node [style=Z phase dot] (52) at (3.25, 0) {$\frac{\pi}{2}$};
		\node [style=Z phase dot] (53) at (4.25, 0) {$\frac{\pi}{2}$};
		\node [style=Z phase dot] (54) at (5.25, 0) {$\frac{\pi}{2}$};
		\node [style=none] (55) at (0.25, 1) {};
		\node [style=none] (56) at (1.25, 1) {};
		\node [style=none] (57) at (2.25, 1) {};
		\node [style=none] (58) at (3.25, 1) {};
		\node [style=none] (59) at (4.25, 1) {};
		\node [style=none] (60) at (5.25, 1) {};
		\node [style=none] (61) at (-11.5, -0.25) {$-2\sqrt{2}i$};
		\node [style=Z dot] (62) at (-6.75, -2.75) {};
		\node [style=hadamard] (63) at (-9.25, -1) {};
		\node [style=hadamard] (64) at (-8.25, -1) {};
		\node [style=hadamard] (65) at (-7.25, -1) {};
		\node [style=hadamard] (66) at (-6.25, -1) {};
		\node [style=hadamard] (67) at (-5.25, -1) {};
		\node [style=hadamard] (68) at (-4.25, -1) {};
		\node [style=Z phase dot] (69) at (-9.25, 0) {$\frac{\pi}{2}$};
		\node [style=Z phase dot] (70) at (-8.25, 0) {$\frac{\pi}{2}$};
		\node [style=Z phase dot] (71) at (-7.25, 0) {$\frac{\pi}{2}$};
		\node [style=Z phase dot] (72) at (-6.25, 0) {$\frac{\pi}{2}$};
		\node [style=Z phase dot] (73) at (-5.25, 0) {$\frac{\pi}{2}$};
		\node [style=Z phase dot] (74) at (-4.25, 0) {$\frac{\pi}{2}$};
		\node [style=none] (75) at (-9.25, 1) {};
		\node [style=none] (76) at (-8.25, 1) {};
		\node [style=none] (77) at (-7.25, 1) {};
		\node [style=none] (78) at (-6.25, 1) {};
		\node [style=none] (79) at (-5.25, 1) {};
		\node [style=none] (80) at (-4.25, 1) {};
		\node [style=none] (81) at (-2, 6.25) {$+2e^{i\pi/4}$};
		\node [style=Z phase dot] (82) at (2.7, 5.5) {-$\frac{\pi}{2}$};
		\node [style=none] (95) at (0.25, 7.25) {};
		\node [style=none] (96) at (1.25, 7.25) {};
		\node [style=none] (97) at (2.25, 7.25) {};
		\node [style=none] (98) at (3.25, 7.25) {};
		\node [style=none] (99) at (4.25, 7.25) {};
		\node [style=none] (100) at (5.25, 7.25) {};
		\node [style=none] (101) at (-11.5, -4.75) {$+8\sqrt{2}i$};
		\node [style=Z dot] (102) at (-9.25, -5.5) {};
		\node [style=Z dot] (103) at (-8.25, -5.5) {};
		\node [style=Z dot] (104) at (-7.25, -5.5) {};
		\node [style=Z dot] (105) at (-6.25, -5.5) {};
		\node [style=Z dot] (106) at (-5.25, -5.5) {};
		\node [style=Z phase dot] (107) at (-4.25, -5.5) {$\pi$};
		\node [style=none] (108) at (-9.25, -4) {};
		\node [style=none] (109) at (-8.25, -4) {};
		\node [style=none] (110) at (-7.25, -4) {};
		\node [style=none] (111) at (-6.25, -4) {};
		\node [style=none] (112) at (-5.25, -4) {};
		\node [style=none] (113) at (-4.25, -4) {};
		\node [style=hadamard] (114) at (-9.25, -4.75) {};
		\node [style=hadamard] (115) at (-8.25, -4.75) {};
		\node [style=hadamard] (116) at (-7.25, -4.75) {};
		\node [style=hadamard] (117) at (-6.25, -4.75) {};
		\node [style=hadamard] (118) at (-5.25, -4.75) {};
		\node [style=none] (119) at (-2, -4.75) {$+8\sqrt{2}i$};
		\node [style=Z dot] (120) at (0, -5.5) {};
		\node [style=Z dot] (121) at (1, -5.5) {};
		\node [style=Z dot] (122) at (3, -5.5) {};
		\node [style=Z dot] (123) at (4, -5.5) {};
		\node [style=Z dot] (124) at (5, -5.5) {};
		\node [style=Z phase dot] (125) at (2, -5.5) {$\pi$};
		\node [style=none] (126) at (0, -4) {};
		\node [style=none] (127) at (1, -4) {};
		\node [style=none] (128) at (3, -4) {};
		\node [style=none] (129) at (4, -4) {};
		\node [style=none] (130) at (5, -4) {};
		\node [style=none] (131) at (2, -4) {};
		\node [style=hadamard] (132) at (0, -4.75) {};
		\node [style=hadamard] (133) at (1, -4.75) {};
		\node [style=hadamard] (134) at (3, -4.75) {};
		\node [style=hadamard] (135) at (4, -4.75) {};
		\node [style=hadamard] (136) at (5, -4.75) {};
	\end{pgfonlayer}
	\begin{pgfonlayer}{edgelayer}
		\draw (2.center) to (1);
		\draw (4.center) to (3);
		\draw (6.center) to (5);
		\draw (8.center) to (7);
		\draw (10.center) to (9);
		\draw (12.center) to (11);
		\draw (16.center) to (15);
		\draw (18.center) to (17);
		\draw (20.center) to (19);
		\draw (22.center) to (21);
		\draw (24.center) to (23);
		\draw (26.center) to (25);
		\draw (29.center) to (28);
		\draw (31.center) to (30);
		\draw (33.center) to (32);
		\draw (35.center) to (34);
		\draw (37.center) to (36);
		\draw (39.center) to (38);
		\draw (43) to (42);
		\draw (44) to (42);
		\draw (45) to (42);
		\draw (46) to (42);
		\draw (47) to (42);
		\draw (48) to (42);
		\draw (49) to (43);
		\draw (50) to (44);
		\draw (51) to (45);
		\draw (52) to (46);
		\draw (53) to (47);
		\draw (54) to (48);
		\draw (55.center) to (49);
		\draw (57.center) to (51);
		\draw (56.center) to (50);
		\draw (58.center) to (52);
		\draw (59.center) to (53);
		\draw (60.center) to (54);
		\draw (63) to (62);
		\draw (64) to (62);
		\draw (65) to (62);
		\draw (66) to (62);
		\draw (67) to (62);
		\draw (68) to (62);
		\draw (69) to (63);
		\draw (70) to (64);
		\draw (71) to (65);
		\draw (72) to (66);
		\draw (73) to (67);
		\draw (74) to (68);
		\draw (75.center) to (69);
		\draw (77.center) to (71);
		\draw (76.center) to (70);
		\draw (78.center) to (72);
		\draw (79.center) to (73);
		\draw (80.center) to (74);
		\draw [in=180, out=-90, looseness=0.75] (95.center) to (82);
		\draw [in=-90, out=165, looseness=0.75] (82) to (96.center);
		\draw [in=135, out=-90, looseness=0.75] (97.center) to (82);
		\draw [in=270, out=45, looseness=0.75] (82) to (98.center);
		\draw [in=15, out=-90, looseness=0.75] (99.center) to (82);
		\draw [in=0, out=-90, looseness=0.75] (100.center) to (82);
		\draw (108.center) to (102);
		\draw (110.center) to (104);
		\draw (109.center) to (103);
		\draw (111.center) to (105);
		\draw (112.center) to (106);
		\draw (113.center) to (107);
		\draw [style=hadamard edge, bend right=75] (102) to (107);
		\draw [style=hadamard edge, bend right=75] (103) to (107);
		\draw [style=hadamard edge, bend right=60] (104) to (107);
		\draw [style=hadamard edge, bend right=45, looseness=1.25] (105) to (107);
		\draw [style=hadamard edge] (106) to (107);
		\draw [style=hadamard edge, bend right=45] (102) to (104);
		\draw [style=hadamard edge, bend right=60] (102) to (105);
		\draw [style=hadamard edge, bend right] (103) to (105);
		\draw [style=hadamard edge, bend right=45, looseness=1.25] (103) to (106);
		\draw [style=hadamard edge, bend right=45] (104) to (106);
		\draw (126.center) to (120);
		\draw (128.center) to (122);
		\draw (127.center) to (121);
		\draw (129.center) to (123);
		\draw (130.center) to (124);
		\draw (131.center) to (125);
		\draw [style=hadamard edge, bend right=75] (120) to (125);
		\draw [style=hadamard edge] (121) to (125);
		\draw [style=hadamard edge] (122) to (125);
		\draw [style=hadamard edge, bend left=45, looseness=0.75] (123) to (125);
		\draw [style=hadamard edge, bend left=75, looseness=1.25] (124) to (125);
		\draw [style=hadamard edge, bend right=60, looseness=1.25] (120) to (122);
		\draw [style=hadamard edge, bend right=90, looseness=1.25] (120) to (123);
		\draw [style=hadamard edge, bend right=60, looseness=1.25] (121) to (123);
		\draw [style=hadamard edge, bend right=90, looseness=1.25] (121) to (124);
		\draw [style=hadamard edge, bend right=60] (122) to (124);
	\end{pgfonlayer}
\end{tikzpicture}
\end{equation}
This decomposition of ZX-diagrams was obtained by finding an explicit local equivalence between each of the stabiliser summands from \cite{bravyi2016trading} to either a product state, a GHZ state, or a more general graph state. Stabiliser states in this form are then readily expressible as ZX-diagrams. The coefficients were adjusted to account for different normalisation conventions.

Now we pick 6 non-Clifford spiders from the diagram and apply the spider-fusion rule (the top-left rule from Figure~\ref{fig:zx-rules}) from right-to-left to isolate, or `unfuse', a $\frac\pi4$ factor from each of these spiders:
\begin{equation}
  \begin{tikzpicture}
	\begin{pgfonlayer}{nodelayer}
		\node [style=Z phase dot] (41) at (-11.25, -2.25) {$\frac{\pi}{2}$};
		\node [style=Z dot] (45) at (-12, 2.5) {};
		\node [style=Z dot] (117) at (-10.25, 2.5) {};
		\node [style=Z phase dot] (119) at (-6.25, 0.25) {$\frac{\pi}{2}$};
		\node [style=Z dot] (123) at (-8.75, 2.5) {};
		\node [style=Z dot] (127) at (-7.75, 2.5) {};
		\node [style=Z phase dot] (143) at (6, -0.75) {$\frac{\pi}{4}$};
		\node [style=Z dot] (147) at (-6.25, 2.5) {};
		\node [style=Z dot] (155) at (-4.75, 2.5) {};
		\node [style=Z dot] (165) at (-3.25, 2.5) {};
		\node [style=Z phase dot] (178) at (8.5, -0.5) {$\frac{\pi}{4}$};
		\node [style=Z dot] (183) at (-1.5, 2.5) {};
		\node [style=Z dot] (186) at (0.5, 2.5) {};
		\node [style=Z phase dot] (188) at (0.75, -1) {$\frac{3\pi}{2}$};
		\node [style=Z dot] (189) at (2.5, 2.5) {};
		\node [style=Z dot] (193) at (3.5, 2.5) {};
		\node [style=Z phase dot] (236) at (14, -3.25) {$\frac{5\pi}{4}$};
		\node [style=Z dot] (245) at (5.25, 2.5) {};
		\node [style=Z dot] (250) at (6.25, 2.5) {};
		\node [style=Z phase dot] (253) at (3.75, -3.5) {$\pi$};
		\node [style=Z dot] (261) at (7.75, 2.5) {};
		\node [style=Z dot] (272) at (9.25, 2.5) {};
		\node [style=Z phase dot] (285) at (16.5, -1.25) {$\frac{7\pi}{4}$};
		\node [style=Z dot] (289) at (11, 2.5) {};
		\node [style=Z dot] (293) at (12, 2.5) {};
		\node [style=Z dot] (328) at (13.5, 2.5) {};
		\node [style=Z dot] (331) at (15.5, 2.5) {};
		\node [style=Z dot] (334) at (17.5, 2.5) {};
		\node [style=Z phase dot] (337) at (15.25, -5) {$\frac{\pi}{4}$};
		\node [style=Z phase dot] (391) at (11, 3.5) {$\frac{\pi}{4}$};
		\node [style=Z phase dot] (392) at (-6.25, 3.5) {$\frac{7\pi}{4}$};
		\node [style=Z phase dot] (397) at (3.5, 3.5) {$\frac{\pi}{4}$};
		\node [style=Z phase dot] (402) at (11.5, -0.5) {$\frac{5\pi}{4}$};
		\node [style=Z phase dot] (407) at (-4.75, 3.5) {$\frac{\pi}{4}$};
		\node [style=Z phase dot] (408) at (0.5, 3.5) {$\frac{7\pi}{4}$};
		\node [style=Z phase dot] (411) at (7.75, 3.5) {$\frac{\pi}{4}$};
		\node [style=Z phase dot] (413) at (13.5, 3.5) {$\frac{\pi}{4}$};
		\node [style=Z phase dot] (416) at (17.5, 3.5) {$\frac{7\pi}{4}$};
		\node [style=Z phase dot] (421) at (2.5, 3.5) {$\frac{\pi}{4}$};
		\node [style=Z phase dot] (424) at (15.5, 3.5) {$\frac{3\pi}{4}$};
		\node [style=Z phase dot] (425) at (-3.25, 3.5) {$\frac{5\pi}{4}$};
		\node [style=Z phase dot] (426) at (5.25, 3.5) {$\frac{7\pi}{4}$};
		\node [style=Z phase dot] (430) at (9.25, 3.5) {$\frac{7\pi}{4}$};
		\node [style=Z phase dot] (434) at (-1.5, 3.5) {$\frac{\pi}{4}$};
		\node [style=Z phase dot] (435) at (-8.75, 3.5) {$\frac{\pi}{4}$};
		\node [style=Z phase dot] (436) at (6.25, 3.5) {$\frac{\pi}{4}$};
		\node [style=Z phase dot] (438) at (-6.5, -3.75) {$\frac{\pi}{2}$};
		\node [style=Z phase dot] (439) at (-10.25, 3.5) {$\frac{7\pi}{4}$};
		\node [style=Z phase dot] (440) at (-12, 3.5) {$\frac{5\pi}{4}$};
		\node [style=Z phase dot] (443) at (12, 3.5) {$\frac{5\pi}{4}$};
		\node [style=Z phase dot] (447) at (-3, -1.5) {$\frac{3\pi}{2}$};
		\node [style=Z phase dot] (454) at (10, -3.75) {$\frac{\pi}{4}$};
		\node [style=Z phase dot] (461) at (-7.75, 3.5) {$\frac{5\pi}{4}$};
		\node [style=Z phase dot] (462) at (-9, -6.5) {$\frac{\pi}{4}$};
		\node [style=Z phase dot] (463) at (-7.75, -6.5) {$\frac{\pi}{4}$};
		\node [style=Z phase dot] (464) at (-6.5, -6.5) {$\frac{\pi}{4}$};
		\node [style=Z phase dot] (465) at (-5.25, -6.5) {$\frac{\pi}{4}$};
		\node [style=Z phase dot] (466) at (-4, -6.5) {$\frac{\pi}{4}$};
		\node [style=Z phase dot] (467) at (-3, -6.5) {$\frac{\pi}{4}$};
	\end{pgfonlayer}
	\begin{pgfonlayer}{edgelayer}
		\draw [style=hadamard edge] (41) to (454);
		\draw [style=hadamard edge] (41) to (447);
		\draw [style=hadamard edge] (41) to (45);
		\draw [style=hadamard edge] (41) to (402);
		\draw [style=hadamard edge] (41) to (117);
		\draw [style=hadamard edge] (41) to (178);
		\draw [style=hadamard edge] (41) to (438);
		\draw [style=hadamard edge] (41) to (188);
		\draw [style=hadamard edge] (41) to (253);
		\draw [style=hadamard edge] (41) to (127);
		\draw [style=hadamard edge] (45) to (438);
		\draw [style=hadamard edge] (45) to (440);
		\draw [style=hadamard edge] (117) to (119);
		\draw [style=hadamard edge] (117) to (438);
		\draw [style=hadamard edge] (117) to (439);
		\draw [style=hadamard edge] (119) to (143);
		\draw [style=hadamard edge] (119) to (337);
		\draw [style=hadamard edge] (119) to (402);
		\draw [style=hadamard edge] (119) to (123);
		\draw [style=hadamard edge] (119) to (447);
		\draw [style=hadamard edge] (119) to (438);
		\draw [style=hadamard edge] (119) to (127);
		\draw [style=hadamard edge] (119) to (253);
		\draw [style=hadamard edge] (123) to (438);
		\draw [style=hadamard edge] (123) to (435);
		\draw [style=hadamard edge] (127) to (461);
		\draw [style=hadamard edge] (143) to (193);
		\draw [style=hadamard edge] (143) to (183);
		\draw [style=hadamard edge] (143) to (337);
		\draw [style=hadamard edge] (143) to (155);
		\draw [style=hadamard edge] (143) to (165);
		\draw [style=hadamard edge] (143) to (438);
		\draw [style=hadamard edge] (143) to (186);
		\draw [style=hadamard edge] (147) to (454);
		\draw [style=hadamard edge] (147) to (392);
		\draw [style=hadamard edge] (147) to (402);
		\draw [style=hadamard edge] (147) to (253);
		\draw [style=hadamard edge] (147) to (447);
		\draw [style=hadamard edge] (155) to (454);
		\draw [style=hadamard edge] (155) to (402);
		\draw [style=hadamard edge] (155) to (407);
		\draw [style=hadamard edge] (155) to (253);
		\draw [style=hadamard edge] (155) to (447);
		\draw [style=hadamard edge] (165) to (454);
		\draw [style=hadamard edge] (165) to (402);
		\draw [style=hadamard edge] (165) to (425);
		\draw [style=hadamard edge] (165) to (253);
		\draw [style=hadamard edge] (178) to (193);
		\draw [style=hadamard edge] (178) to (454);
		\draw [style=hadamard edge] (178) to (236);
		\draw [style=hadamard edge] (178) to (253);
		\draw [style=hadamard edge] (178) to (438);
		\draw [style=hadamard edge] (178) to (183);
		\draw [style=hadamard edge] (178) to (189);
		\draw [style=hadamard edge] (183) to (434);
		\draw [style=hadamard edge] (183) to (188);
		\draw [style=hadamard edge] (183) to (447);
		\draw [style=hadamard edge] (186) to (188);
		\draw [style=hadamard edge] (186) to (408);
		\draw [style=hadamard edge] (186) to (447);
		\draw [style=hadamard edge] (188) to (454);
		\draw [style=hadamard edge] (188) to (236);
		\draw [style=hadamard edge] (188) to (253);
		\draw [style=hadamard edge] (188) to (438);
		\draw [style=hadamard edge] (188) to (189);
		\draw [style=hadamard edge] (189) to (421);
		\draw [style=hadamard edge] (193) to (397);
		\draw [style=hadamard edge] (193) to (447);
		\draw [style=hadamard edge] (236) to (261);
		\draw [style=hadamard edge] (236) to (454);
		\draw [style=hadamard edge] (236) to (328);
		\draw [style=hadamard edge] (236) to (334);
		\draw [style=hadamard edge] (236) to (402);
		\draw [style=hadamard edge] (236) to (285);
		\draw [style=hadamard edge] (236) to (293);
		\draw [style=hadamard edge] (236) to (289);
		\draw [style=hadamard edge] (236) to (245);
		\draw [style=hadamard edge] (236) to (250);
		\draw [style=hadamard edge] (236) to (447);
		\draw [style=hadamard edge] (245) to (402);
		\draw [style=hadamard edge] (245) to (426);
		\draw [style=hadamard edge] (245) to (253);
		\draw [style=hadamard edge] (250) to (454);
		\draw [style=hadamard edge] (250) to (436);
		\draw [style=hadamard edge] (250) to (253);
		\draw [style=hadamard edge] (253) to (272);
		\draw [style=hadamard edge] (253) to (402);
		\draw [style=hadamard edge] (253) to (285);
		\draw [style=hadamard edge] (253) to (293);
		\draw [style=hadamard edge] (253) to (438);
		\draw [style=hadamard edge] (253) to (447);
		\draw [style=hadamard edge] (253) to (454);
		\draw [style=hadamard edge] (253) to (331);
		\draw [style=hadamard edge] (253) to (334);
		\draw [style=hadamard edge] (253) to (337);
		\draw [style=hadamard edge] (261) to (454);
		\draw [style=hadamard edge] (261) to (402);
		\draw [style=hadamard edge] (261) to (411);
		\draw [style=hadamard edge] (272) to (285);
		\draw [style=hadamard edge] (272) to (430);
		\draw [style=hadamard edge] (285) to (454);
		\draw [style=hadamard edge] (285) to (289);
		\draw [style=hadamard edge] (285) to (293);
		\draw [style=hadamard edge] (289) to (391);
		\draw [style=hadamard edge] (289) to (402);
		\draw [style=hadamard edge] (293) to (402);
		\draw [style=hadamard edge] (293) to (443);
		\draw [style=hadamard edge] (328) to (413);
		\draw [style=hadamard edge] (328) to (337);
		\draw [style=hadamard edge] (328) to (402);
		\draw [style=hadamard edge] (331) to (424);
		\draw [style=hadamard edge] (331) to (337);
		\draw [style=hadamard edge] (334) to (416);
		\draw [style=hadamard edge] (334) to (337);
		\draw [style=hadamard edge] (334) to (402);
		\draw [style=hadamard edge] (337) to (438);
		\draw [style=hadamard edge] (337) to (447);
		\draw [style=hadamard edge] (402) to (438);
		\draw [style=hadamard edge] (438) to (447);
		\draw [style=hadamard edge] (447) to (454);
		\draw (41) to (462);
		\draw (463) to (119);
		\draw (464) to (438);
		\draw (465) to (447);
		\draw (466) to (188);
		\draw (467) to (253);
	\end{pgfonlayer}
\end{tikzpicture}
\end{equation}

We then apply the decomposition~\eqref{eq:magic-state-decomposition} to these 6 spiders with the $\frac\pi4$ phase, resulting in 7 diagrams where these non-Clifford spiders have been replaced by some Clifford subdiagram.
We then perform the necessary spider fusions to reduce each of the diagrams to graph-like form.
For each of the diagrams we again apply the simplification algorithm of Section~\ref{sec:simplify-ZX} so that all the new Clifford spiders are removed, and often some  additional non-Clifford spiders are combined as well. Now for each of the diagrams we pick 6 non-Clifford spiders to apply the decomposition to and we repeat the procedure. If there aren't 6 non-Clifford spiders then the diagram is small enough that we can directly calculate the scalar it represents, or alternatively, we can apply the pairwise stabiliser decomposition to reduce it further.

Once every Clifford spider has been removed from a diagram, we are only left with the scalar factor that arose from application of the simplification rules in section~\ref{sec:simplify-ZX}, which gives us the amplitude of a single term in the decomposition. We compute the overall amplitude by summing these numbers together.

Because we are simplifying the diagrams after each decomposition, we will sometimes see cancellations of non-Clifford phases. This is the primary benefit of our method, as the number of non-Clifford phases in the computation is what leads to the exponential cost of the simulation. However, even if we don't find any cancellations there is an asymptotic benefit to our method.

Suppose none of the non-Clifford phases cancel. Then after applying the magic state decomposition~\eqref{eq:magic-state-decomposition} we require a constant number of rewrites to simplify the diagram. As the size of the diagrams is $O(t)$, a rewrite requires at most $O(t^2)$ graph operations (searching for rewrites to apply requires $O(t)$ operations and hence can be ignored). There will be $2^{\alpha t}$ diagrams, all of size at most $O(t)$. Hence, the number of graph operations required is $O(2^{\alpha t}t^2)$.
This compares favourably to the $O(2^{\alpha t}t^3)$ found in~\cite{bravyi2016trading}. Note that the exponential factor is of course of much higher import, and our method performs better in practice not because of this asymptotic improvement, but because we have to do fewer decompositions due to the reduction in the number of non-Clifford phases.

Calculating marginal probabilities is very similar to the calculation of single amplitudes. However, rather than starting by translating $\<x_1\cdots x_n|U|0\cdots0\>$ to a ZX-diagram, we translate the last expression in equation~\eqref{eq:marginal} to obtain the following, `doubled' ZX-diagram:
\begin{equation}
  P(x_1 \cdots x_k) \ =\  \begin{tikzpicture}
	\begin{pgfonlayer}{nodelayer}
		\node [style=X dot] (32) at (-9.5, 2.25) {};
		\node [style=X phase dot] (33) at (-1.25, 2.25) {$\ x_1\pi~$};
		\node [style=X dot] (34) at (-9.5, 0.75) {};
		\node [style=X phase dot] (35) at (-1.25, 0.75) {$\ x_k\pi~$};
		\node [style=X dot] (36) at (-9.5, -0.75) {};
		\node [style=X dot] (53) at (-9.5, -2.75) {};
		\node [style=none] (55) at (-7.5, 2.75) {};
		\node [style=none] (56) at (-7.5, -3.25) {};
		\node [style=none] (57) at (-3.5, -3.25) {};
		\node [style=none] (58) at (-3.5, 2.75) {};
		\node [style=none] (59) at (-7.5, 2.25) {};
		\node [style=none] (60) at (-7.5, 0.75) {};
		\node [style=none] (61) at (-7.5, -0.75) {};
		\node [style=none] (63) at (-7.5, -2.75) {};
		\node [style=none] (64) at (-3.5, -2.75) {};
		\node [style=none] (66) at (-3.5, -0.75) {};
		\node [style=none] (67) at (-3.5, 0.75) {};
		\node [style=none] (68) at (-3.5, 2.25) {};
		\node [style=none] (69) at (-5.5, -0.125) {$U$};
		\node [style=none, rotate=90] (168) at (-8.25, -1.75) {$\cdots$};
		\node [style=none, rotate=90] (169) at (-2.25, -1.75) {$\cdots$};
		\node [style=none, rotate=90] (170) at (-2.25, 1.5) {$\cdots$};
		\node [style=X dot] (171) at (9.5, 2.25) {};
		\node [style=X phase dot] (172) at (1.25, 2.25) {$\ x_1\pi~$};
		\node [style=X dot] (173) at (9.5, 0.75) {};
		\node [style=X phase dot] (174) at (1.25, 0.75) {$\ x_k\pi~$};
		\node [style=X dot] (175) at (9.5, -0.75) {};
		\node [style=X dot] (177) at (9.5, -2.75) {};
		\node [style=none] (179) at (7.5, 2.75) {};
		\node [style=none] (180) at (7.5, -3.25) {};
		\node [style=none] (181) at (3.5, -3.25) {};
		\node [style=none] (182) at (3.5, 2.75) {};
		\node [style=none] (183) at (7.5, 2.25) {};
		\node [style=none] (184) at (7.5, 0.75) {};
		\node [style=none] (185) at (7.5, -0.75) {};
		\node [style=none] (186) at (7.5, -2.75) {};
		\node [style=none] (187) at (3.5, -2.75) {};
		\node [style=none] (188) at (3.5, -0.75) {};
		\node [style=none] (189) at (3.5, 0.75) {};
		\node [style=none] (190) at (3.5, 2.25) {};
		\node [style=none] (191) at (5.5, -0.125) {$U^\dagger$};
		\node [style=none, rotate=90] (192) at (8.25, -1.75) {$\cdots$};
		\node [style=none, rotate=90] (193) at (2.25, -1.75) {$\cdots$};
		\node [style=none, rotate=90] (194) at (2.25, 1.5) {$\cdots$};
		\node [style=none, rotate=90] (195) at (-8.25, 1.5) {$\cdots$};
		\node [style=none, rotate=90] (196) at (8.25, 1.5) {$\cdots$};
	\end{pgfonlayer}
	\begin{pgfonlayer}{edgelayer}
		\draw (55.center) to (58.center);
		\draw (58.center) to (57.center);
		\draw (57.center) to (56.center);
		\draw (56.center) to (55.center);
		\draw (68.center) to (33);
		\draw (53) to (63.center);
		\draw (36) to (61.center);
		\draw (60.center) to (34);
		\draw (32) to (59.center);
		\draw (67.center) to (35);
		\draw (179.center) to (182.center);
		\draw (182.center) to (181.center);
		\draw (181.center) to (180.center);
		\draw (180.center) to (179.center);
		\draw (190.center) to (172);
		\draw (177) to (186.center);
		\draw (175) to (185.center);
		\draw (184.center) to (173);
		\draw (171) to (183.center);
		\draw (189.center) to (174);
		\draw (66.center) to (188.center);
		\draw (187.center) to (64.center);
	\end{pgfonlayer}
\end{tikzpicture}
\end{equation}
This is a special case of the technique sometimes referred to as the \textit{CPM-construction}~\cite{selinger2007dagger} or simply \textit{doubling}~\cite{CKbook} in the diagrammatic/categorical quantum mechanics literature.


Since we still represent a marginal probability as a single ZX-diagram, we can compute it using the exact same method we described for computing single amplitudes. 
However, note that we now have both $U$ and $U^\dagger$, each of which contains $t$ non-Clifford gates so that this diagram contains $2t$ non-Clifford gates.
Naively this means that calculating a marginal probability will take (exponentially) longer than calculating a single amplitude. However, as $U$ connects directly to $U^\dagger$ on some of the wires, many of the gates on both sides can cancel out. This is in fact what we observe in practice, so that calculating a marginal probability does not seem to take much longer.

\subsection{Implementation}\label{sec:imp}

We implemented the stabiliser rank decomposition in \texttt{quizx}~\cite{quizx}, a port into the Rust programming language of the \texttt{PyZX} library for representing and rewriting quantum circuits and ZX-diagrams~\cite{pyzx}. Instructions for downloading and building the tool, as well as reproducing the data in the next section, are available at:

\begin{center}
  \url{https://github.com/Quantomatic/quizx/blob/stabrank-v1/stabrank.md}
\end{center}

The algorithm is implemented as described in the previous section, where ZX-diagrams are represented as sparse undirected graphs. All scalar factors are stored as elements of the ring $\mathbb D[e^{i\pi/4}]$, where $\mathbb D$ is the ring of dyadic rational numbers. In other words, we store scalars as five integer parameters $(k,a,b,c,d)$ representing the complex number $\frac{1}{2^k} (a + b e^{i\pi/4} + c i + d e^{-i \pi/4})$. Using this representation, as opposed to floating point complex numbers, we can store the values of any scalars arising from Clifford+T circuits exactly~\cite{giles2013exact}.
Probabilities arise as positive real numbers in this ring, which all have the form $\frac{1}{2^k} (x + y e^{i\pi/4} + y e^{-i \pi/4}) = \frac{1}{2^k}(x + y\sqrt{2})$.

To compute amplitudes, we first do a partial decomposition of a ZX-diagram to a small fixed depth $d$ (in our case $d=3$), then evaluate the amplitude of each of the resulting $K \leq 7^d$ diagrams in parallel. For each of these diagrams, the remaining decomposition is done in a depth-first fashion, so at most $7^d \cdot \lceil \frac t 6 \rceil$ diagrams are kept in memory at any given time. This enables us to make use of a multicore system without generating too much memory overhead.

\section{Results}\label{sec:results}

To assess the effectiveness of our approach, we performed classical simulation for two families of circuits: (i) random Clifford+T circuits arising from exponentiated Pauli unitaries and (ii) random instances of the hidden shift algorithm. The latter is exactly the family of circuits considered in~\cite{bravyiImprovedClassicalSimulation2016}.

\subsection{Random Clifford+T circuits}

\begin{figure}
  \centering

  \begin{minipage}{0.48\textwidth}
    \centering

    \includegraphics[height=4.8cm]{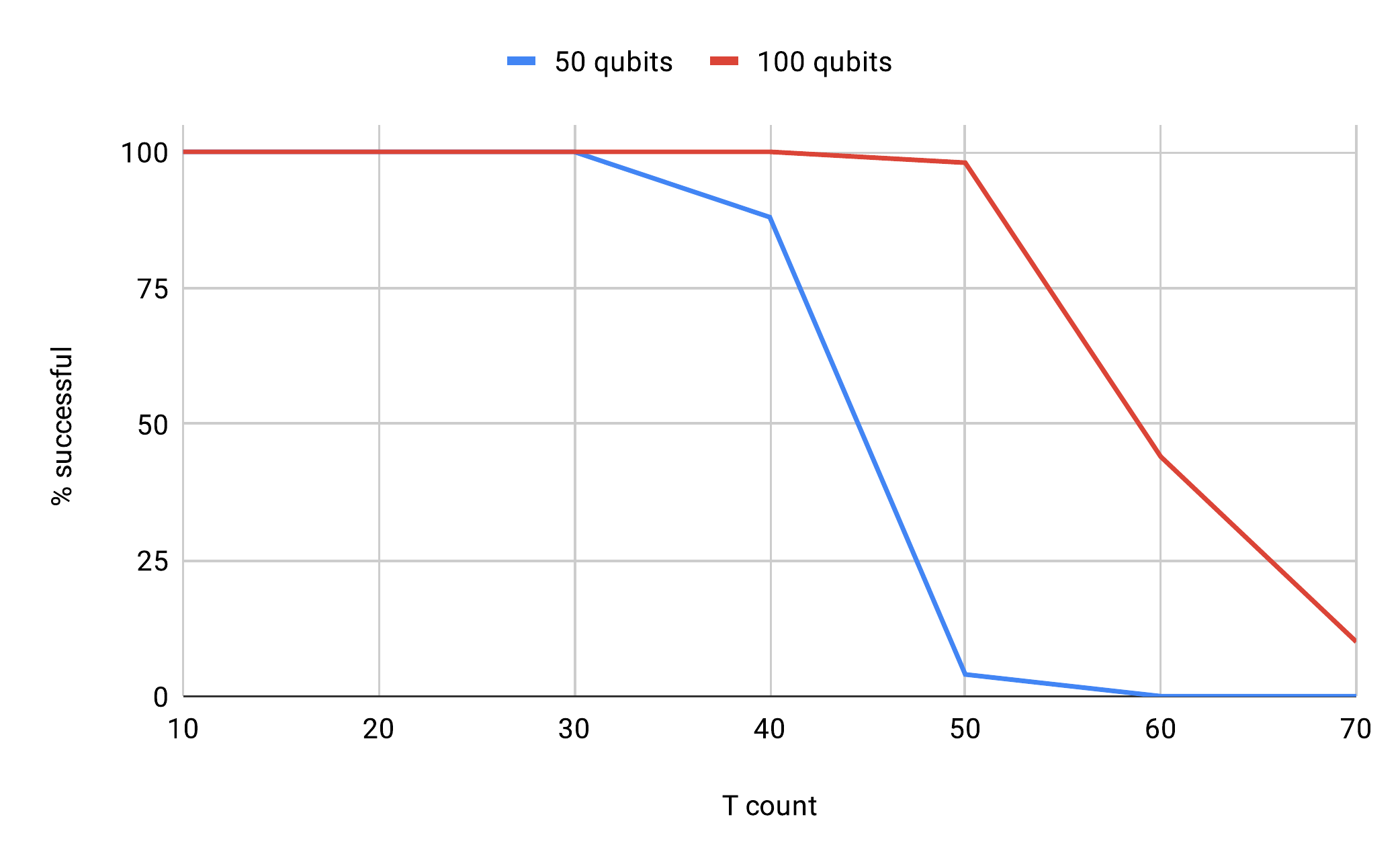}
  \end{minipage}

  \caption{\label{fig:random-success} Percentage of random 50- and 100-qubit circuits of a given T-count that were sucessfully sampled in under 5 minutes. For each T-count 50 random circuits were generated.}
\end{figure}

An \textit{exponentiated Pauli unitary} is a unitary of the form $U_{P,\alpha} := \exp(-i \frac\alpha 2 P)$ for $P$ an element of the $N$-qubit Pauli group $P \propto P_1 \otimes \ldots \otimes P_n$ for $P_i \in \{ I, X, Y, Z \}$. We say $U_{P,\alpha}$ has \textit{weight} $k$ if $P$ contains at most $k$ non-identity Pauli operators. Operators of this form arise naturally in various contexts, including generic representations of Clifford+T circuits~\cite{gosset2014tcount,Litinski2019gameofsurfacecodes} and Hamiltonian simulation via trotterisation~\cite{phaseGadgetSynth,cowtan2020generic}.

We generate random Clifford+T circuits by fixing a number of exponentiated Paulis which will in turn fix the T count. For each exponentiated Pauli $U_{P,\alpha}$, we choose a random Pauli string $P$ with a weight $k$ between some fixed minimum and maximum weights, as well as a random odd integer multiple of $\frac \pi 4$ for $\alpha$. In this case, $U_{P,\alpha}$ can always be implemented with exactly 1 T gate, $2(k-1)$ CNOT gates, and local Cliffords~\cite{phaseGadgetSynth}.

For our experiments, we generated random 50 and 100 qubit circuits with minimum weight 2 and maximum weight 4. Depths (a.k.a. T counts) range from 10 to 70. For each circuit, we use strong simulation to produce one full sample according to the output distribution of the circuit. Namely, we compute the marginal probability $P(q_1 = 0)$ of obtaining outcome $0$ on the first qubit, then fix $b_0 \in \{ 0, 1 \}$ probabilistically according to $P(q_1 = 0)$. We then apply $\bra{0}$ or $\bra{1}$ to the first output qubit according to the value of $b_1$, compute the probability $P(q_2 = 0 \,|\, q_1 = b_1)$ in order to select $b_2$, and so on until we obtain the full bitstring $b_1, \ldots, b_N$ as well as an exact expression for the probability $P(q_1 = b_1, \ldots, q_N = b_N)$.

As this is a heuristic method, it is difficult to estimate the runtime, apart from a (typically infeasible) upper bound of $O(N \cdot 7^{2t/6}t^2)$ for $N$ qubits and $t$ T gates, which comes from applying the na\"ive BSS decomposition to the doubled circuit. 
So to demonstrate the feasability of our method, we did numerical experiments. 
We ran all circuits with a timeout of 5 minutes on a dedicated computation server (24-core Intel Xeon E5-2667, 2.90GHz) and computed the percentage that were successfully simulated before the timeout (Fig.~\ref{fig:random-success}).

We found that for random circuits the simulation performed well up to a certain T-count and then the success-rate fell off rapidly, as one would expect from the exponential complexity of the method. For random 50-qubit circuits, we could simulate in less than 5 minutes all circuits with T-counts up to 30, the majority (88\%) of circuits with T-count 40, 4\% of circuits with T-count 50, and no circuits with higher T-counts. What might seem initially counter-intuitive is that 100-qubit circuits actually perform better: all circuits of T-count 40 or less are simulated within 5 minutes, along with all but one of the T-count 50 circuits, 44\% of the T-count 60 circuits, and 10\% of the T-count 70 circuits.

To understand this effect, it is worth noting that these circuits with large numbers of qubits are very sparse. For 50-qubit circuits of depth 70, each qubit participates in $4.2$ Pauli exponentials on average. For the 100-qubit circuits, each qubit will participate in half as many. Empirically, it seems to be the case that sparser circuits enable the ZX-calculus simplifier to eliminate more T gates, which has a drastic effect on simulation times.

\begin{figure}
  \centering

  \begin{minipage}{0.48\textwidth}
    \footnotesize\it\centering Reduction vs. na\"ive BSS decomposition

    \includegraphics[height=4.8cm]{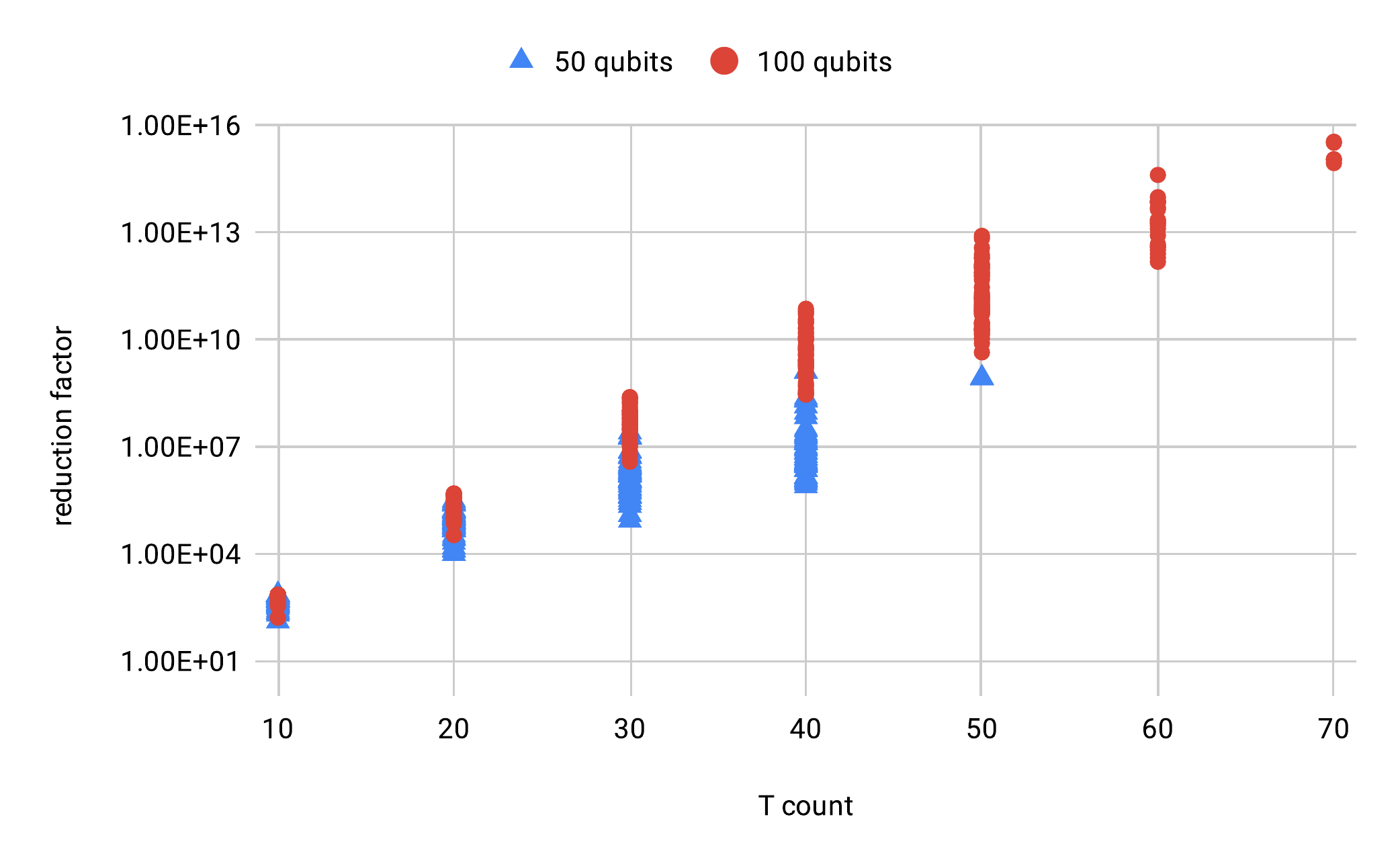}
  \end{minipage}
  \begin{minipage}{0.48\textwidth}
    \footnotesize\it\centering  Reduction vs. simplified BSS decomposition

    \includegraphics[height=4.8cm]{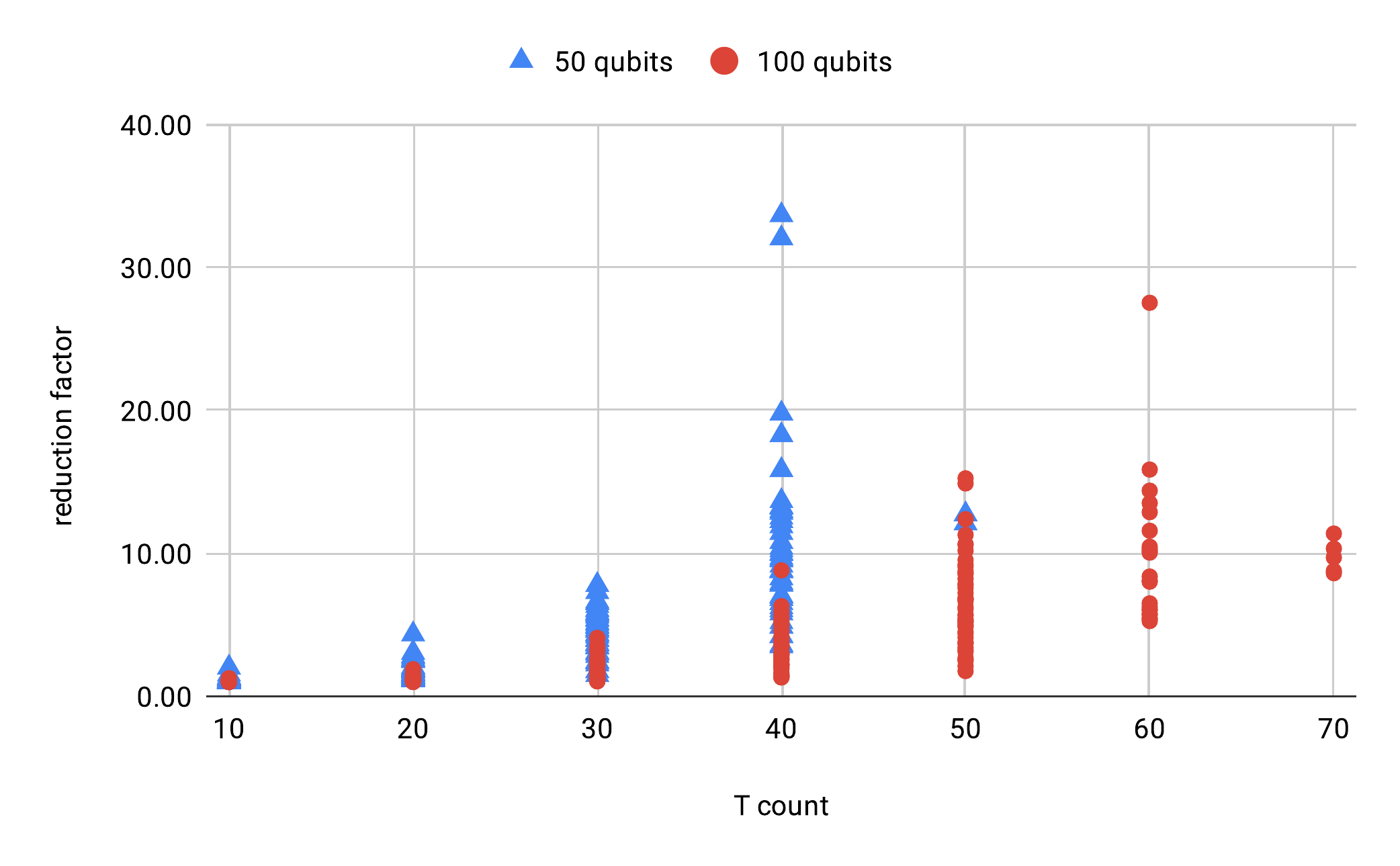}
 \end{minipage}

  \caption{\label{fig:random-improve} Reduction in term count on random 50- and 100-qubit circuits of varying T-counts. For each circuit simulated, the `reduction factor' indicates the number of stabiliser terms needed using na\"ive BSS decomposition (left) or BSS decomposition after a single ZX-simplification (right), divided by the number of stabiliser terms used by our technique.}
\end{figure}

In all cases, the ZX-calculus simplified decomposition produced much lower numbers of stabiliser terms than na\"ive BSS decompositions of the doubled circuits (Fig.~\ref{fig:random-improve}, left). For example, strong simulation with the na\"ive BSS decomposition on a 50-qubit circuit with 40 T gates requires approximately $50 \cdot 7^{2\cdot 40/6} \approx 9 \times 10^{12}$ terms whereas the ZX-calculus simplified decomposition of the random circuits produced an average of $2.6\times 10^6$ terms, an improvement of 6 orders of magnitude. In the case of higher T-counts, we see in Fig.~\ref{fig:random-improve} (left) improvements of up to 15 orders of magnitude.

It is natural to ask how much of this effect is due to the interleaved ZX-calculus simplifications, and how much comes from the initial simplification of the doubled circuit, where many gates may cancel and the T-count might otherwise be reduced before we begin. To see the effect of interleaving the simplification, we can compare our technique to doing a full BSS decomposition not on the original doubled circuit, but on one that has already been optimised for T-count. For random circuits, we still get improved performance, but it is not nearly as striking. In Fig.~\ref{fig:random-improve} (right), we see about a 10X reduction on average in the number of stabiliser terms if we interleave simplifications, with a peak of about 35X. In this regime of essentially unstructured circuits, it may therefore be more effective to do T-count reduction in advance and then hand off the resulting (post-selected) circuit to a fast tableau simulator.

We will see in the next section that for structured circuits, this story can be very different.



\subsection{Hidden shift circuits}

We will now apply our simulation procedure to random 50-qubit instances of the hidden shift algorithm. This is the same family of circuits considered in~\cite{Bravyi2019simulationofquantum} and~\cite{bravyiImprovedClassicalSimulation2016}. A detailed description of the generation of these circuits can be found in the supplemental material of the latter.\footnote{Note the published version contains an error which has been corrected in the arXiv preprint (v3)~\cite{bravyi2016improvedarxiv}.} 
Briefly, each circuit contains two classical oracles which together `hide' a bitstring in their relationship with each other, which can be recovered deterministically from a single query to each oracle using a simple quantum algorithm~\cite{rotteler2010hiddenshift}.

There are several notable properties for this family of circuits, which make them a good testbed. The first is that we can define the classical oracle such that the hidden shift is known, so we can always check that the simulator gets the right answer. The second is that, since the hidden shift algorithm gives a deterministic outcome, any joint distribution is a product of the single-qubit marginals. Hence, it is only necessary to compute single-qubit marginals to fully simulate the circuit.
Finally, by varying the number of CCZ gates which appear in the classical oracles, we can precisely control the number of T gates which occur in the resulting circuit.

The hidden shift circuits are initially generated in the gate set H, Z, CZ, CCZ. To produce a ZX-diagram for this circuit, we must translate each of these gates into ZX-diagrams. The ZX-diagrams for H, Z, and CZ were already introduced in Section~\ref{sec:zx}. For CCZ, we have (at least) two useful encodings into a ZX-diagram: one that has 7 non-Clifford spiders and one that has 4 non-Clifford spiders. The 7-spider version is the following:
\begin{equation}\label{eq:7t-ccz}
  \begin{tikzpicture}
	\begin{pgfonlayer}{nodelayer}
		\node [style=none] (0) at (-1, 0) {};
		\node [style=none] (1) at (-1, 1.5) {};
		\node [style=none] (2) at (-1, -1.5) {};
		\node [style=cnot ctrl] (3) at (0, 0) {};
		\node [style=cnot ctrl] (4) at (0, 1.5) {};
		\node [style=cnot ctrl] (5) at (0, -1.5) {};
		\node [style=none] (6) at (1, 0) {};
		\node [style=none] (7) at (1, 1.5) {};
		\node [style=none] (8) at (1, -1.5) {};
	\end{pgfonlayer}
	\begin{pgfonlayer}{edgelayer}
		\draw (4) to (3);
		\draw (3) to (5);
		\draw (0.center) to (6.center);
		\draw (1.center) to (7.center);
		\draw (2.center) to (8.center);
	\end{pgfonlayer}
\end{tikzpicture} \ =\ 
  \sqrt{2}^5\ 
  \begin{tikzpicture}
	\begin{pgfonlayer}{nodelayer}
		\node [style=Z phase dot] (3) at (0, -2) {$\frac\pi4$};
		\node [style=none] (4) at (-2, 2) {};
		\node [style=none] (5) at (-2, 0) {};
		\node [style=none] (6) at (-2, -2) {};
		\node [style=none] (7) at (2, 2) {};
		\node [style=none] (8) at (2, 0) {};
		\node [style=none] (9) at (2, -2) {};
		\node [style=Z dot] (10) at (1, -1) {};
		\node [style=Z dot] (11) at (-1, 1) {};
		\node [style=Z dot] (12) at (1, 1) {};
		\node [style=Z phase dot] (13) at (-2, 1) {-$\frac\pi4$};
		\node [style=Z phase dot] (14) at (2, -1) {-$\frac\pi4$};
		\node [style=Z phase dot] (15) at (2, 1) {$\frac\pi4$};
		\node [style=Z phase dot] (16) at (0, 0) {$\frac\pi4$};
		\node [style=Z phase dot] (17) at (0, 2) {$\frac\pi4$};
		\node [style=Z dot] (18) at (-1, -1) {};
		\node [style=Z phase dot] (19) at (-2, -1) {-$\frac\pi4$};
	\end{pgfonlayer}
	\begin{pgfonlayer}{edgelayer}
		\draw (6.center) to (3);
		\draw (3) to (9.center);
		\draw [style=hadamard edge] (10) to (14);
		\draw [style=hadamard edge] (11) to (13);
		\draw [style=hadamard edge] (12) to (15);
		\draw (5.center) to (16);
		\draw (16) to (8.center);
		\draw [style=hadamard edge] (16) to (12);
		\draw [style=hadamard edge] (16) to (10);
		\draw (4.center) to (17);
		\draw (17) to (7.center);
		\draw [style=hadamard edge] (17) to (12);
		\draw [style=hadamard edge] (17) to (11);
		\draw [style=hadamard edge] (18) to (19);
		\draw [style=hadamard edge] (12) to (3);
		\draw [style=hadamard edge] (3) to (10);
		\draw [style=hadamard edge] (11) to (16);
		\draw [style=hadamard edge] (18) to (3);
		\draw [style=hadamard edge] (18) to (17);
	\end{pgfonlayer}
\end{tikzpicture}
\end{equation}
It arises from simplifying the `textbook' presentation of CCZ or Toffoli in terms of CNOT and T gates (see e.g.~\cite{NielsenChuang}, Section 4.3) or from the graphical Fourier transform~\cite{GraphicalFourier2019,vandewetering2020zxcalculus}. The 4-spider version is the following:
\begin{equation}\label{eq:4t-ccz}
  \begin{tikzpicture}
	\begin{pgfonlayer}{nodelayer}
		\node [style=none] (0) at (-1, 0) {};
		\node [style=none] (1) at (-1, 1.5) {};
		\node [style=none] (2) at (-1, -1.5) {};
		\node [style=cnot ctrl] (3) at (0, 0) {};
		\node [style=cnot ctrl] (4) at (0, 1.5) {};
		\node [style=cnot ctrl] (5) at (0, -1.5) {};
		\node [style=none] (6) at (1, 0) {};
		\node [style=none] (7) at (1, 1.5) {};
		\node [style=none] (8) at (1, -1.5) {};
	\end{pgfonlayer}
	\begin{pgfonlayer}{edgelayer}
		\draw (4) to (3);
		\draw (3) to (5);
		\draw (0.center) to (6.center);
		\draw (1.center) to (7.center);
		\draw (2.center) to (8.center);
	\end{pgfonlayer}
\end{tikzpicture} \ =\ 
  4e^{i\frac\pi4}\ 
  \begin{tikzpicture}
	\begin{pgfonlayer}{nodelayer}
		\node [style=Z dot] (0) at (0, 3) {};
		\node [style=Z dot] (1) at (0, 1) {};
		\node [style=Z phase dot] (2) at (0, -1) {-$\frac\pi4$};
		\node [style=Z phase dot] (3) at (0, -2.25) {-$\frac\pi2$};
		\node [style=none] (4) at (-2, 3) {};
		\node [style=none] (5) at (-2, 1) {};
		\node [style=none] (6) at (-2, -2.25) {};
		\node [style=none] (7) at (2, 3) {};
		\node [style=none] (8) at (2, 1) {};
		\node [style=none] (9) at (2, -2.25) {};
		\node [style=Z dot] (10) at (1, 0) {};
		\node [style=Z dot] (11) at (1, 2) {};
		\node [style=Z dot] (12) at (-1, 0) {};
		\node [style=Z phase dot] (13) at (2, 2) {-$\frac\pi4$};
		\node [style=Z phase dot] (14) at (2, 0) {-$\frac\pi4$};
		\node [style=Z phase dot] (15) at (-2, 0) {$\frac\pi4$};
	\end{pgfonlayer}
	\begin{pgfonlayer}{edgelayer}
		\draw [style=hadamard edge] (2) to (3);
		\draw (4.center) to (0);
		\draw (0) to (7.center);
		\draw (5.center) to (1);
		\draw (1) to (8.center);
		\draw (6.center) to (3);
		\draw (3) to (9.center);
		\draw [style=hadamard edge] (10) to (14);
		\draw [style=hadamard edge] (11) to (13);
		\draw [style=hadamard edge] (12) to (15);
		\draw [style=hadamard edge] (0) to (12);
		\draw [style=hadamard edge] (1) to (12);
		\draw [style=hadamard edge] (2) to (12);
		\draw [style=hadamard edge] (0) to (11);
		\draw [style=hadamard edge] (2) to (11);
		\draw [style=hadamard edge] (1) to (10);
		\draw [style=hadamard edge] (2) to (10);
	\end{pgfonlayer}
\end{tikzpicture}
\end{equation}
It can be derived by simplifying the (postselected version of the) CCZ gadget introduced by Jones~\cite{jones2013low}. Either can be verified by concrete calculation.

While the second decomposition results in lower initial T-counts, the first enables the ZX-calculus simplifier to find many more rewrites. For example, the simplification strategy from Section~\ref{sec:simplify-ZX} is able to cancel out the composition of two copies of \eqref{eq:7t-ccz}, but not~\eqref{eq:4t-ccz}. For this reason, we choose the first decomposition for translating hidden shift circuits into ZX-diagrams.

\begin{figure}
  \centering
  \begin{minipage}{0.48\textwidth}
    \centering

    \includegraphics[height=4.8cm]{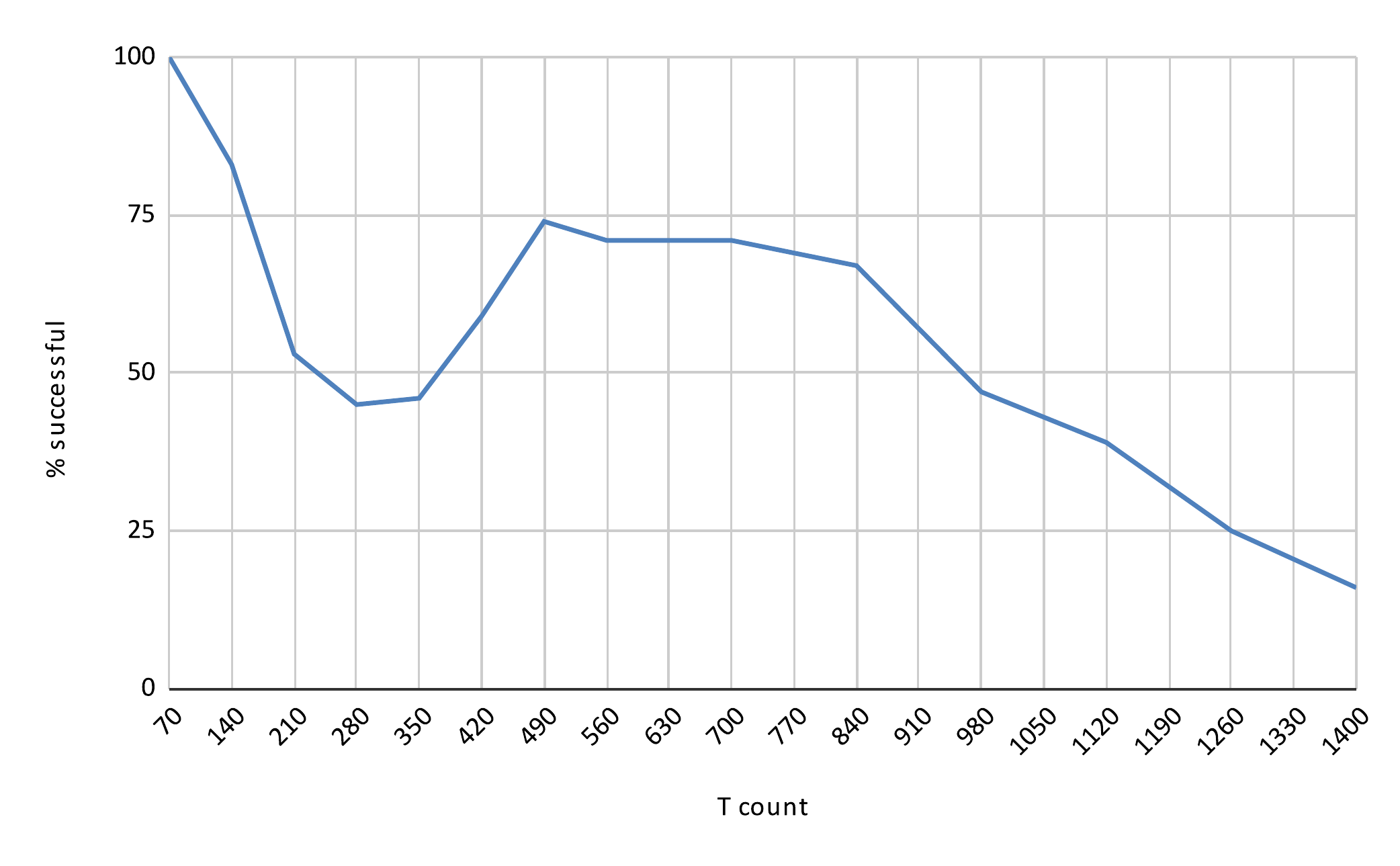}
  \end{minipage}

  \caption{\label{fig:hidden-shift-success} Percentage of 100 random 50-qubit hidden shift circuits successfully sampled in less than 5 minutes.}
\end{figure}

\begin{figure}
  \centering

  \begin{minipage}{0.48\textwidth}
    \footnotesize\it\centering Reduction vs. na\"ive BSS decomposition
    
    \includegraphics[height=4.8cm]{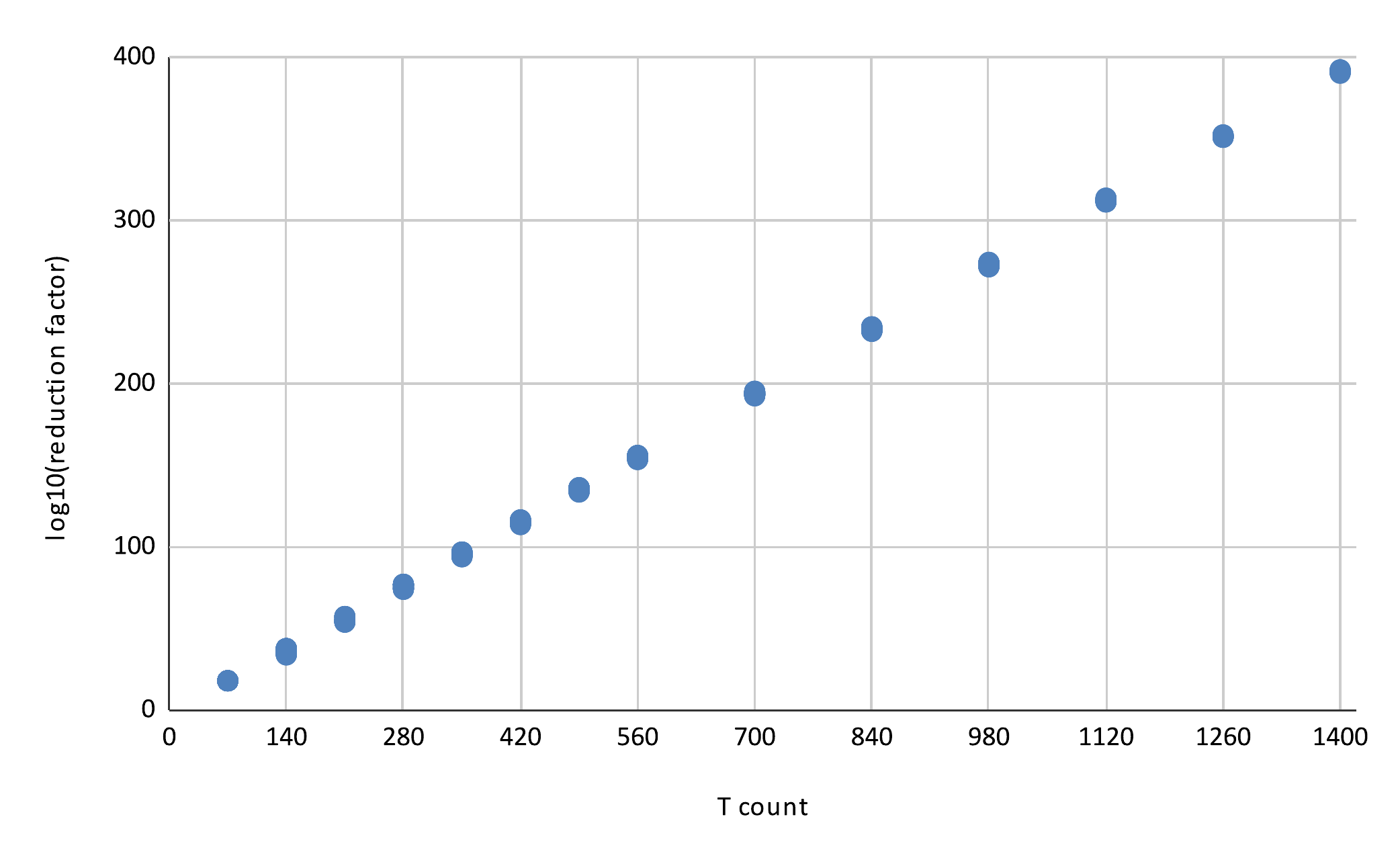}
  \end{minipage}
  \begin{minipage}{0.48\textwidth}
    \footnotesize\it\centering Reduction vs. simplified BSS decomposition

    \includegraphics[height=4.8cm]{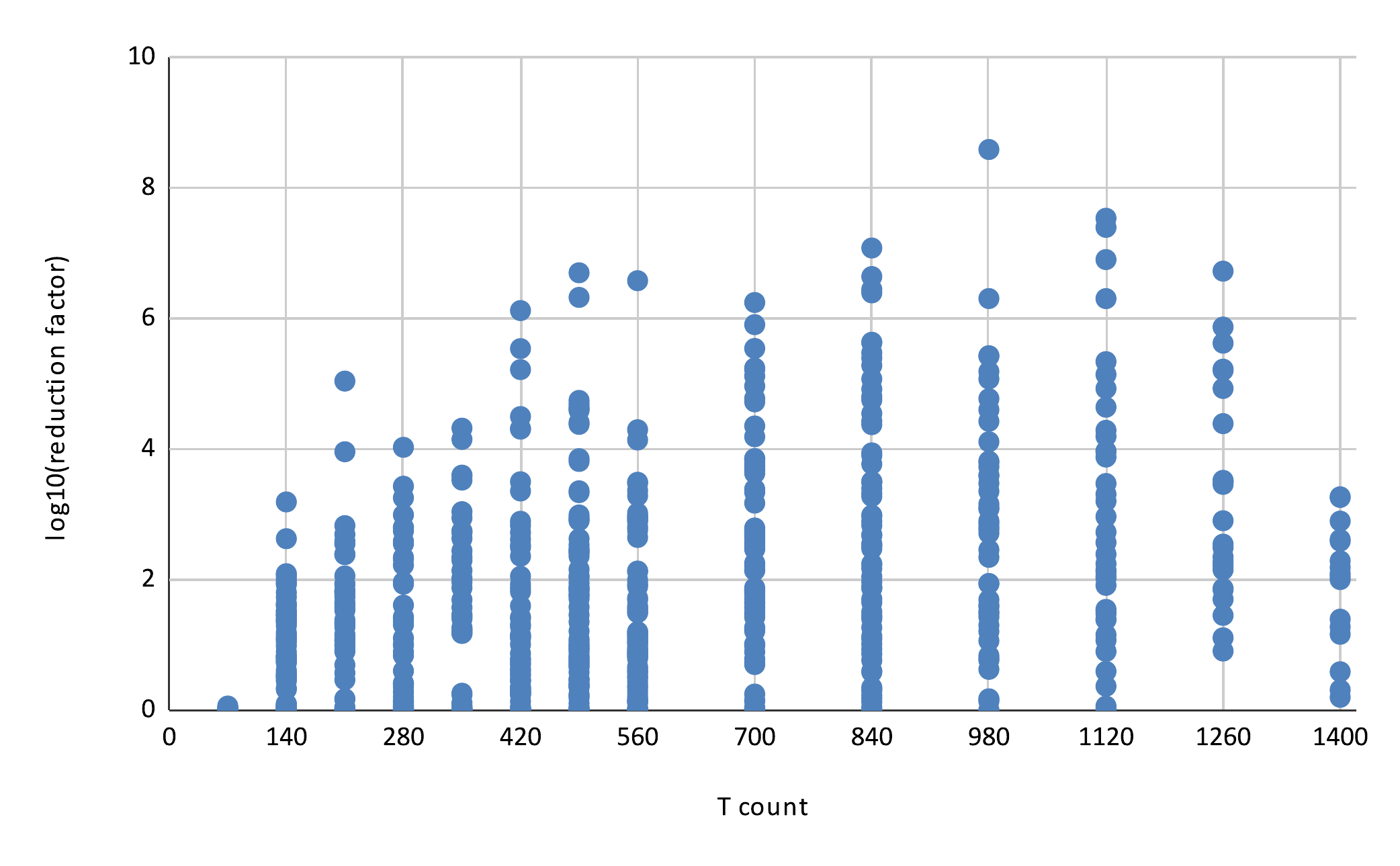}
  \end{minipage}

  \caption{\label{fig:hidden-shift-improve} Plots of reduction in term count on 50-qubit hidden shift circuits vs. na\"ive BSS decomposition (left) and BSS decomposition after a single ZX-simplification (right).}
\end{figure}

As before, we fix a 5-minute timeout for producing a single full sample of each circuit and compute the percentage of circuits that were successful (Fig.~\ref{fig:hidden-shift-success}). For each given CCZ count we generated 100 random, 50-qubit hidden shift instances and computed each' single-qubit marginals of all the qubits. We vary the CCZ counts of the circuitx from 10 to 80 in increments of 10, then 100 to 200 in increments of 20. Reported T-counts in Fig.~\ref{fig:hidden-shift-success} and Fig.~\ref{fig:hidden-shift-improve} are then 7 * CCZ count.

As mentioned above, this is a deterministic algorithm, so we fix the outcome on the $i$-th qubit according to this marginal, i.e. $b_i = 0$ if $P(q_i = 0) = 1$ and $b_i = 1$ otherwise. We then verified for all successfully-simulated circuits that this indeed matches the hidden shift between the two classical oracles.

For a CCZ count of 10 (T-count 70) we observe 100\% success, then some fluctuation in the success rate until a CCZ count of 120 (T-count 840), followed by a much more gradual drop in success rate as compared with the random circuits. We were successful in simulating the largest family of circuits (200 CCZ / 1400 T) 17\% of the time.

Interestingly, unlike the random circuit case, interleaving ZX simplification still produced substantial benefits vs. a single round of T-count optimisation (Fig.~\ref{fig:hidden-shift-improve}), yielding many reductions in stabiliser terms of more than 6 orders of magnitude.

\section{Conclusion and Future Work}\label{sec:conclusion}


We have shown how we can use ZX-diagram simplification to improve quantum simulation methods based on stabiliser decompositions. 
We successfully simulated random 50- and 100-qubit Pauli exponential circuits with up to 70 T gates, as well as 50-qubit hidden shift circuits with up to 1400 T gates.
For the former class of circuits, most of the benefit of our method comes from the initial T-count optimisation. This is particularly apparent for the 100-qubit circuits which are quite sparse, and hence where a lot more non-Clifford gates act on trivial states or effects. Our results hence make clear that for stabiliser decomposition based simulation methods it is important to cancel as many non-Clifford gates in the original circuit.
For the hidden shift circuits there is also an enormous benefit from cancelling T gates in the first optimisation step, but subsequent simplifications on the non-unitary diagrams also contribute significantly to the reduced simulation time of our method.

There are many areas where our method can potentially be improved.
A natural next step is to incorporate the improved magic state decompositions presented in Ref.~\cite{qassim2021improved} into our method. This will involve finding convenient ZX-diagram presentations for the stabiliser states in these decompositions and running some further experiments.


Another way our method can be improved is by utilising a more powerful ZX-calculus simplification strategy. Our strategy is that of Ref.~\cite{kissinger2019tcount}, which is specifically intended for unitary circuits, and as such does not include rewrites that only apply in a non-unitary context. In particular, we could use the \emph{supplementarity} rule of Ref.~\cite{supplementarity} to cancel additional non-Clifford phases.
We could also include additional simplification strategies to reduce the T-count of the diagram. While the strategies of Refs.~\cite{amy2016t,heyfron2018efficient} are probably too slow to apply to thousands of diagrams in parallel, the heuristics of Ref.~\cite{deBeaudrap2020Techniques} might be fast enough to apply to the first couple of layers of diagrams. For this to work best, we will need to find the best trade-off between the two (theoretically) exponential tasks of full T-count minimisation and stabiliser decomposition. It may be the case that, with a more powerful simplifier, we can switch to the more efficient T-count 4 representation~\eqref{eq:4t-ccz} for CCZ gates without sacrificing nice simplification behaviour in the resulting ZX-diagrams.

Another approach to consider would be to directly represent the CCZ gates using the \emph{H-boxes} of the \emph{ZH-calculus}~\cite{backens2018zhcalculus,zhcompleteness2020}. A CCZ gate has a stabiliser rank of 2, so this could potentially lead to a significant reduction in the number of terms needed for simplification. Instead of the simplification strategy of Section~\ref{sec:simplify-ZX} based on local complementation, we could then use the strategy of Ref.~\cite{Lemonnier2020hypergraph} that uses \emph{hyper-local-complementation}. However, this method has a tendency to increase the number of H-boxes, so care must be taken for this method to actually lead to a reduction in the number of terms required for simulation.

We have only considered exact strong simulation in this paper. In particular, our exact calculation of marginals doubles the T-count right at the beginning. This can be avoided by instead doing norm estimation using random stabiliser states sampled from a projective $t$-design~\cite{Bravyi2019simulationofquantum}. However, in preliminary experiments, we found this method to be prohibitively slow at obtaining enough samples to approximate the norm to high precision. Part of the problem here is that, for calculating many independent amplitudes, ZX-calculus simplification seems to be quite a bit slower than a fast tableau simulator. Perhaps this could be improved by mixing ZX and tableau-based simulation or finding better ways to share some of the simplification work across all the samples.

Finally, it would be interesting to see whether techniques based on the stabiliser extent~\cite{Bravyi2019simulationofquantum}, rather than the stabiliser rank, can be accelerated by the ZX-calculus. In this approach, summands are sampled according to the relative magnitudes of their coefficients rather than computed exhaustively. The amount of samples needed to obtain a good approximation of the state then depends on the stabiliser extent, which effectively measures how much negativity is present in the list of coefficients. One possible avenue for incorporating ZX-calculus simplification is to adopt a hybrid approach. For example, one could perform the ZX-diagram decomposition we have described here up to a fixed depth, then approximate the value of each summand using random sampling based on the stabiliser extent.

In this work, we achieved substantial improvements on prior work by combining known techniques for classical simulation and optimisation of quantum circuits in a relatively simple way. The broader lesson here is that we should not treat classical simulation and optimisation as two separate problems, but rather as two complementary ways to extract information about the behaviour of a quantum circuit, which can be used together fruitfully.



\paragraph{Acknowledgments:} AK is supported by AFOSR grant FA2386-18-1-4028. JvdW is funded by a NWO Rubicon personal fellowship. The authors would like to thank Niel de Beaudrap, Mark Howard, Earl Campbell, and Padraic Calpin for useful discussions related to the stabiliser decompositions in~\cite{Bravyi2019simulationofquantum}.

\bibliographystyle{plain}
\bibliography{main}

\end{document}